\documentclass[journal]{IEEEtran}
\usepackage[T1]{fontenc}
\usepackage{amsmath,amssymb,bm}
\usepackage{booktabs,multirow,array,longtable}
\usepackage{graphicx}
\usepackage{tikz}
\usetikzlibrary{arrows.meta,decorations.pathreplacing,positioning}
\usepackage{cite}
\usepackage[hidelinks]{hyperref}
\usepackage{orcidlink}
\graphicspath{{./}}

\newcommand{\RMSE}{\mathrm{RMSE}}
\newcommand{\cRMSE}{\mathrm{cRMSE}}
\newcommand{\MAE}{\mathrm{MAE}}
\usepackage{dblfloatfix}
\usepackage{xcolor}
\usepackage{makecell}
\newcommand{\fitcol}[1]{\resizebox{\ifdim\width>\columnwidth\columnwidth\else\width\fi}{!}{#1}}

\begin{document}
\bstctlcite{IEEEexample:BSTcontrol}

\title{Hardware-Aware Design and Layout Selection in Quantum Reservoir Computing for Insurance Claims Forecasting}

\author{
Renáta Rusnáková\,\orcidlink{0009-0004-3966-2201},
Eugen Šlapak\,\orcidlink{0000-0003-1629-7876},
Gaurav Kumar Pandey\,\orcidlink{0000-0001-9959-2891},~\IEEEmembership{Member,~IEEE},
and Juraj Gazda\,\orcidlink{0000-0002-7334-9540}%
\thanks{All authors are with the Department of Computers and Informatics, Faculty of Electrical Engineering and Informatics, Technical University of Košice, Košice, Slovakia. 
Corresponding author: Renáta Rusnáková
(e-mail: renata.rusnakova@tuke.sk).

This work has been submitted to the IEEE for possible publication. Copyright may be transferred without notice, after which this version may no longer be accessible.}}

\markboth{Hardware-aware quantum reservoir forecasting of insurance claims}{Hardware-aware quantum reservoir forecasting of insurance claims}

\maketitle

\begin{abstract}
    Quantum reservoir computing (QRC) processes temporal information through quantum dynamics, but it remains unclear whether small QRC models can forecast real data competitively with classical models and whether their predictions survive the transfer from ideal simulation to noisy hardware. We address both questions for a six-qubit gate-based reservoir that forecasts next-day motor-insurance claim counts. Five reservoir candidates were selected and frozen using hardware-native interaction graphs, Trotterized Ising dynamics, memory and measurement-robustness criteria, circuit cost, and calibrated-noise screening. On 232 held-out test days, the ideal candidates reached root-mean-square errors (RMSEs) of 5.66--6.12 claims/day, compared with 5.36 for ridge regression. For four of the five, the difference from ridge was not significant under a 95\% block bootstrap, and their accuracy was comparable with classical recurrent and reservoir baselines. On the 156-qubit IBM Kingston processor, we ran 63 evaluations in 21 paired rounds, comparing hardware-aware, random-native, and Qiskit optimization level 1 layouts. Hardware-aware placement gave lower RMSE than Qiskit in every round, with a maximum transfer distortion of 7.59 claims/day (11.84 for Qiskit), and avoided the severe transfer failures seen for some random layouts. Most of the hardware distortion appeared as a systematic forecast shift, which a short hardware probe on training inputs detected before the test runs. Longer replay windows raised circuit cost without a consistent gain in forecasting accuracy or hardware transfer. Small QRC models can thus reach competitive accuracy on this task, and hardware-aware layout selection makes their transfer to IBM Kingston more consistent.

\end{abstract}


\section{Introduction}
    \IEEEPARstart{C}{urrent} quantum processors have tens to hundreds of qubits, gate errors around $10^{-3}$, coherence times of a few hundred microseconds, and calibration data that change from hour to hour \cite{Preskill2018NISQ,Bharti2022NISQ,Klimov2018Fluctuations}. Algorithms that require long coherent evolution or deep, trained circuits are difficult to run reliably on such machines. Reservoir computing is attractive under these constraints because the dynamical system is fixed and only a linear readout is trained \cite{Jaeger2004ESN,Lukosevicius2009RC}. In quantum reservoir computing (QRC), the reservoir is a driven many-body quantum system, and the readout acts on measured expectation values \cite{Fujii2017QRC,Mujal2021Review}. Because model fitting does not require gradient-based optimization of quantum-circuit parameters, the barren-plateau and gradient-estimation issues associated with variational quantum algorithms do not arise. The difficulty is instead shifted from simulation-based training to QPU transfer. The readout is fitted to features computed in simulation, whereas the processor returns observables altered by decoherence, gate and readout errors, and finite sampling. Whether a QRC forecaster preserves its simulated behavior on hardware therefore depends both on the reservoir design and on the physical qubits on which it is executed \cite{Chen2020Noisy,Murali2019NoiseAdaptive,Tannu2019NotAll,Niu2020HardwareAware,Nation2023Mapomatic}.

    For a fixed reservoir and readout, any hardware-induced change in the measured observables propagates directly into the predictions. Forecasting accuracy and transfer distortion must therefore be measured separately. We study reservoir design and physical-qubit placement together but evaluate them with separate forecasting and transfer metrics.
    
\subsection{Related Work}\label{sec:related_work}

    \begin{table*}[t]
    \centering
    \caption{Closest QRC Studies Compared With This Work}
    \label{tab:related}
    \footnotesize
    \setlength{\tabcolsep}{4pt}
        \begin{tabular}{@{}lllll@{}}
        \toprule
        Work & Encoding & Reservoir; readout & Hardware (qubits) & Task \\
        \midrule
        Fujii and Nakajima \cite{Fujii2017QRC} & State injection & Transverse-field Ising; linear & Numerical simulation ($5$--$7$) & Temporal benchmarks \\
        Kutvonen \emph{et al.} \cite{Kutvonen2020Optimizing} & State injection & Fully connected transverse-field Ising; linear & Numerical simulation ($\leq 7$) & Stock prices \\
        Chen \emph{et al.} \cite{Chen2020Noisy} & Gate encoding & Gate-based; linear & IBM Boeblingen/Ourense/Vigo ($4$--$10$) & Temporal benchmarks \\
        Suzuki \emph{et al.} \cite{Suzuki2022Natural} & Gate encoding & Natural noisy dynamics; linear & IBM Melbourne/Toronto ($8$--$12$) & NARMA, classification \\
        Wudarski \emph{et al.} \cite{Wudarski2023Hybrid} & Projection + angles & Fixed entangling; ridge & OQC Lucy ($6$--$8$) & Chaotic systems \\
        Hamhoum \emph{et al.} \cite{Hamhoum2025Multivariate} & Angle encoding & Trotterized transverse-field Ising; ridge & IBM Heron R2 ($6$--$15$) & Lorenz-63, ENSO \\
        Connerty \emph{et al.} \cite{Connerty2026Hardware} & Angle encoding & Gate-based network; elastic net & IBM Marrakesh ($12$) & Lorenz system \\
        Tamturk \emph{et al.} \cite{Tamturk2026Insurance} & Angle encoding & QRC features fed to classical boosting & Qiskit simulator ($3$--$5$) & Insurance claim severity \\
        \midrule
        \textbf{This work} & Angle encoding & Trotterized transverse-field Ising + $Y$; ridge & IBM Kingston ($6$) & Insurance claim counts \\
        \bottomrule
        \end{tabular}
    \end{table*}

\subsubsection{Forecasting Insurance Claims}
    Insurance claim counts are traditionally forecast with Poisson and negative-binomial regression, where portfolio exposure can be incorporated explicitly through an offset or covariate~\cite{Nelder1972GLM,Cameron2013Count,Wuthrich2023Actuarial}. More flexible approaches include structural time-series models, tree ensembles and neural networks~\cite{Harvey1989Claims,Henckaerts2021Boosting}. Recent comparisons on automobile claim-frequency data show that machine-learning models can outperform standard Poisson and negative-binomial (NB2) models, although the best method depends on the data structure. Alomair~\cite{Alomair2024Count}, for example, found support-vector regression to outperform the tested count-regression and neural models on a zero-inflated claim-frequency dataset. 

    Quantum methods have only recently entered insurance modeling. Tamturk \emph{et al.}~\cite{Tamturk2026Insurance} combined QRC-generated features with classical boosting for motor-insurance claim-severity prediction and reported improvements over classical ridge, XGBoost and CatBoost variants, but the quantum component was evaluated on a noise-free simulator rather than a quantum processor. To our knowledge, no previous study has evaluated QRC for temporal insurance claim-count forecasting on real quantum hardware.  

\subsubsection{Classical Reservoir Computing and Recurrent Models}
    Classical reservoirs, particularly echo state networks (ESNs), are the closest classical analogue of QRC because their recurrent dynamics remain fixed and only a linear readout is trained \cite{Jaeger2004ESN,Lukosevicius2009RC}. Their temporal processing is commonly characterized by linear memory capacity, which measures how well past inputs can be reconstructed from the current reservoir state~\cite{Jaeger2002STM}. Dambre \emph{et al.}~\cite{Dambre2012Capacity} extended this idea to information-processing capacity, showing that a reservoir distributes its finite feature capacity between linear memory and nonlinear transformations of present and past inputs. Memory is therefore a central computational property of a reservoir, and in our design we explicitly measure delayed-input memory capacity and use it as one of the selection criteria. The broader Dambre capacity decomposition is also applied in the analysis phase to the frozen candidates to distinguish linear recall from nonlinear processing.

    Physical reservoir computing generalizes the same principle by replacing the artificial recurrent network with the dynamics of a physical substrate~\cite{Tanaka2019Physical}. QRC follows this idea in the quantum domain, where the fixed dynamical system is a quantum evolution, and the measured observables form the feature space used by the classical readout. This contrasts with recurrent neural networks (RNNs) and gated recurrent units (GRUs), whose recurrent dynamics are learned directly from data~\cite{Elman1990RNN,Cho2014GRU}. The comparison between these model families therefore separates three different ways of generating temporal features: fixed classical dynamics, learned recurrent dynamics, and fixed quantum dynamics.

\subsubsection{Quantum Reservoir Computing}
    QRC has been implemented in several forms. Some approaches use fixed gate-based reservoirs rather than an explicit Hamiltonian evolution. Wudarski \emph{et al.}~\cite{Wudarski2023Hybrid} encode inputs through single-qubit rotations, apply fixed entangling layers, and train a classical ridge readout on measured expectation values. Ahmed \emph{et al.}~\cite{Ahmed2024RecurrenceFree} similarly use fixed entangling circuits, while part of the temporal memory is carried by a classical leaky integrator rather than by the quantum system itself. Such hybrid designs show that useful reservoir features can be generated without reproducing a physical many-body Hamiltonian.
   
    A second line of work treats the quantum system itself as the reservoir and lets temporal information be processed by its internal dynamics. In Hamiltonian QRC, particularly implementations based on the transverse-field Ising model, classical inputs are encoded into quantum states or rotation angles, the system evolves under fixed interacting dynamics, and expectation values of selected observables form the feature vector for a linear readout~\cite{Fujii2017QRC,Kutvonen2020Optimizing,MartinezPena2021Phase}. On gate-based processors, this evolution can be implemented through Trotterized circuits, bringing the interaction graph, circuit depth, and measured observables directly into the hardware design.

    Several studies have demonstrated QRC on superconducting quantum processors. Chen \emph{et al.}~\cite{Chen2020Noisy} showed that temporal processing remains viable on noisy IBM hardware when the linear readout is trained directly on hardware-generated reservoir features. Suzuki \emph{et al.}~\cite{Suzuki2022Natural} went further by exploiting the natural noise of a superconducting device as part of the reservoir dynamics itself. More recent work has shown that superconducting QRC can be used for forecasting tasks, including chaotic dynamics on NISQ hardware and multivariate time-series prediction with rewind-based reservoirs on IBM Heron processors \cite{Connerty2026Hardware,Hamhoum2025Multivariate}. Beyond superconducting platforms, large-scale reservoir learning has also been demonstrated on analog neutral-atom hardware~\cite{Kornjaca2024Analog}. Table~\ref{tab:related} summarizes the QRC studies most closely related to the present work.

    Moving QRC from simulation to hardware introduces a second design problem: the logical reservoir must be mapped onto physical qubits whose gate errors, readout errors and coherence properties are non-uniform and time dependent. Noise-adaptive and variability-aware mapping methods therefore use calibration information to avoid weak qubits and couplers~\cite{Murali2019NoiseAdaptive,Tannu2019NotAll,Niu2020HardwareAware}, while approaches such as mapomatic rank candidate layouts by predicted circuit fidelity~\cite{Nation2023Mapomatic}. When the logical interaction graph already embeds natively, the problem reduces largely to choosing the physical placement without routing~\cite{Steinberg2022Assignment}, which is also the setting considered here.

    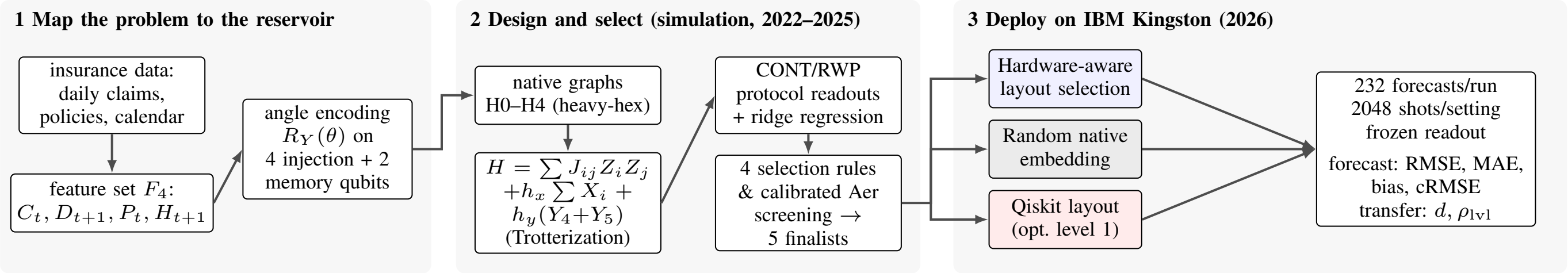
\begin{figure*}[t]
    \centering
    \resizebox{\textwidth}{!}{
        \begin{tikzpicture}[
          font=\scriptsize,
          box/.style={draw, rounded corners=1.5pt, align=center, minimum height=7mm, inner sep=2pt, fill=white},
          stage/.style={draw=black!25, rounded corners=3pt, fill=black!3},
          arr/.style={-{Latex[length=1.6mm]}, thick, black!70},
          lab/.style={font=\scriptsize\bfseries, anchor=north west}]
        \fill[black!3, rounded corners=3pt] (0,0) rectangle (5.2,3.3);
        \fill[black!3, rounded corners=3pt] (5.5,0) rectangle (11.1,3.3);
        \fill[black!3, rounded corners=3pt] (11.5,0) rectangle (18.9,3.3);
        \node[lab] at (0.05,3.28) {1 Map the problem to the reservoir};
        \node[lab] at (5.55,3.28) {2 Design and select (simulation, 2022--2025)};
        \node[lab] at (11.55,3.28) {3 Deploy on IBM Kingston (2026)};
        \node[box, text width=21mm] (data) at (1.35,2.15) {insurance data: daily claims, policies, calendar};
        \node[box, text width=23mm] (feat) at (1.35,0.85) {feature set $F_4$:\\ $C_t,D_{t+1},P_t,H_{t+1}$};
        \node[box, text width=19mm] (enc) at (3.95,1.5) {angle encoding\\ $R_Y(\theta)$ on\\ 4 injection + 2 memory qubits};
        \draw[arr] (data) -- (feat);
        \draw[arr] (feat.east) -- (enc.west);
        \node[box, text width=21mm] (topo) at (6.85,2.15) {native graphs\\ H0--H4 (heavy-hex)};
        \node[box, text width=21mm, align=center] (ham) at (6.85,0.85)
        {$H=\sum J_{ij}Z_iZ_j$\\
        $+h_x\sum X_i+h_y(Y_4{+}Y_5)$\\
        (Trotterization)};
        \node[box, text width=21mm] (rules) at (9.75,2.15) {CONT/RWP protocol readouts + ridge regression};
        \node[box, text width=21mm] (aer) at (9.75,0.85) {4 selection rules \& calibrated Aer\\ screening $\to$ 5 finalists};
        \draw[arr] (enc.east) -- ++(0.35,0) |- (topo.west);
        \draw[arr] (topo) -- (ham);
        \draw[arr] (ham.east) -- (rules.west);
        \draw[arr] (rules) -- (aer);
        \node[box, text width=17mm, fill=blue!6] (ha) at (12.85,2.35) {Hardware-aware layout selection};
        \node[box, text width=17mm, fill=black!8] (rnd) at (12.85,1.5) {Random native\\ embedding};
        \node[box, text width=17mm, fill=red!8] (qk) at (12.85,0.65) {Qiskit layout\\ (opt.\ level 1)};
        \node[box, text width=25mm] (qpu) at (17.2,1.5)
        {232 forecasts/run\\
        2048 shots/setting\\
        frozen readout\\[1mm]
        forecast: RMSE, MAE,\\ bias, cRMSE\\
        transfer: $d,\rho_{\rm lvl}$};
        \draw[arr] (aer.east) -- ++(0.35,0) |- (rnd.west);
        \draw[arr] (aer.east) ++(0.35,0) |- (ha.west);
        \draw[arr] (aer.east) ++(0.35,0) |- (qk.west);
        \draw[arr] (ha.east) -- (qpu.west);
        \draw[arr] (rnd.east) -- (qpu.west);
        \draw[arr] (qk.east) -- (qpu.west);
        \end{tikzpicture}
        }
    \caption{High-level workflow from forecasting inputs to reservoir design, candidate selection, and QPU evaluation under three layout policies.}
    \label{fig:workflow}
    \end{figure*}

\subsection{Motivation and Contributions of the Work}
    Existing QRC studies provide little evidence on temporal insurance claim-count forecasting on real quantum hardware or on how such forecasts survive simulation-to-QPU transfer. We therefore study next-day insurance claim-count forecasting. This is a standard actuarial time-series problem in which claim frequency is modeled relative to portfolio exposure and calendar effects \cite{Harvey1989Claims,Wuthrich2023Actuarial}. Our dataset contains four and a half years of daily motor third-party-liability property-damage claims from a commercial insurer, together with the number of active policies and calendar information. During the observation period, the insured portfolio grows substantially, resulting in evaluation-period exposure levels beyond those observed during training. The data are partitioned chronologically into training, validation, and held-out test periods, with all feature selection, model tuning, and candidate selection performed without access to the test targets. We ask two research questions:

    \paragraph{Forecasting performance}
        Can the QRC architecture forecast insurance claims competitively with classical statistical, recurrent, and reservoir models when all models use the same inputs and are frozen before testing?

    \paragraph{Simulation-to-QPU transfer}
        Can the architecture be systematically designed and placed on real NISQ hardware to ensure reliable transfer of its forecasts from ideal simulation to the quantum processor?

    To address the first question, we evaluate next-day insurance claim-count forecasting on the IBM Kingston processor and compare QRC against a range of classical approaches, including ridge regression, Poisson and NB2 generalized linear models (GLMs) with exposure modeling, ESNs, RNNs, GRUs, and XGBoost with squared-error and Poisson objectives.

    To systematically investigate the second research question, we develop a problem-to-processor workflow comprising three \textit{stages}, as illustrated in Fig.~\ref{fig:workflow}. In \textit{Stage~1}, the insurance forecasting problem is reduced to the four-input feature set $F_4$, and the inputs are encoded into a six-qubit reservoir containing four injection and two memory qubits.  
    \textit{Stage~2} performs logical and hardware-aware reservoir design using only training and validation data. The six-qubit reservoir evolves under a Trotterized transverse-field Ising Hamiltonian with an additional memory-local $Y$ field. The design includes selecting processor-native interaction graphs (five heavy-hex-compatible topologies), Hamiltonian parameters, temporal protocols, observable sets, and the ridge readout. Both continuous evolution and the rewind protocol (RWP) of Mujal \emph{et al.}~\cite{Mujal2023Measurements} are considered, with RWP reconstructing the temporal context before measurement. Single-qubit and correlator observables provide the features for the classical ridge readout. Five reservoir candidates are subsequently selected using four predefined selection rules and calibrated noisy-simulation screening, and frozen before test evaluation.    
    In \textit{Stage~3}, the five frozen reservoir candidates are evaluated on the held-out test set using IBM Kingston. Each candidate is executed under three physical-layout policies: hardware-aware (HA), random-native, and Qiskit level~1. The hardware-aware policy selects physical qubits based on their ability to preserve ideal reservoir predictions, as estimated using training-only QPU probes, rather than relying solely on hardware error rates or circuit-level fidelity estimates. Unlike approaches that delegate physical mapping to the compiler or retrain the classical readout using hardware-generated features, our methodology retains the frozen forecaster and compares repeated executions across the three layout policies. Performance is assessed using conventional forecasting metrics and transfer distortion, allowing us to quantify both predictive accuracy and the preservation of ideal forecasting behavior on real hardware.    
    
    The two research questions are evaluated separately throughout the study. A reservoir achieving high forecasting accuracy in ideal simulation does not necessarily preserve its predictions when executed on noisy quantum hardware. Conversely, hardware-induced perturbations may occasionally reduce forecasting error through bias compensation without indicating improved simulation-to-QPU transfer. The main contributions of this work are as follows.
    \begin{itemize}
        \item \emph{A reproducible hardware-aware QRC workflow.}
        We provide a fully reproducible pipeline, including the selection rules, intermediate candidate sets, calibration-aware screening, and frozen final configurations. The implementation and recorded outputs are available in the public repository.\footnote{\url{https://github.com/rusnakrenata/qrc_insurance}}
    
        \item \emph{Competitive forecasting against classical baselines.}
        The ideal QRC candidates achieve forecasting accuracy comparable with the strongest classical baselines, although ridge remains the best model on the held-out test period. The main QRC limitation is systematic underprediction as portfolio exposure moves beyond the training range; after removing this level shift, the reservoirs reproduce day-to-day variation competitively.
    
        \item \emph{A repeated hardware comparison of physical-layout policies.}
        We evaluate five frozen reservoir candidates across $21$ paired hardware rounds, yielding $63$ complete QPU evaluations under three physical-layout policies (Fig.~\ref{fig:workflow}). HA achieves lower forecasting RMSE than Qiskit in all $21$ rounds and limits maximum transfer distortion to $7.59$ claims/day, compared with $11.84$ claims/day for Qiskit. Although random-native placement occasionally achieves lower RMSE than HA, it exhibits greater variability, including four severe transfer failures with distortions of $13.0$--$35.3$ claims/day associated with degraded or non-functional couplers.
    
        \item \emph{A pre-deployment diagnosis of QPU transfer error.}
        A short QPU probe using only training inputs identifies systematic prediction-level shifts that subsequently appear in test forecasts. Observable-level analysis further reveals how contraction and offsets in measured reservoir features contribute to these shifts.
    \end{itemize}

    The results show that QRC can be competitive when both the logical reservoir architecture and the physical qubit layout are chosen carefully. On our insurance data, ridge benefits from linear extrapolation to the growing portfolio, while the QRC remains comparable with the ESN, GRU, vanilla RNN, and XGBoost baselines.

    \textit{The organization of the paper is as follows.} Section~\ref{sec:dataset} states the forecasting problem, data, objectives and metrics. Section~\ref{sec:qrm} defines the reservoir. Section~\ref{sec:5candidates} describes how the hardware candidates were designed and selected, and Section~\ref{sec:execution} provides the hardware execution protocol and an overview of the classical baselines. Section~\ref{sec:results} reports results for both objectives, and Section~\ref{sec:discussion} discusses the findings and limitations. Section~\ref{sec:conclusion} concludes the paper.

\section{Dataset, Problem Formulation and Metrics}\label{sec:dataset}
\subsection{Dataset and Predictor Selection}\label{sec:predictors}    
     The dataset was provided by a commercial insurer in the form of daily portfolio records. It contains claim counts by type (total, property damage, injury, and loss of profit), the number of active policies, policy-age counts and proportions, the number and proportion of policies held by legal entities, and the mean vehicle age. Calendar variables include the day of the week, weekend and public-holiday indicators, and working-day status. The dataset was complete, with no missing dates, missing values, or blank entries. It was divided chronologically by target date into training (2022-01 to 2024-12, 1,095 observations), validation (2025-01 to 2025-12, 365 observations), and held-out test (2026-01-01 to 2026-08-20, 232 observations) sets, with each observation pairing the inputs of day $t$ with the claims of day $t+1$. The test period ends on August 20, 2026, corresponding to the last available observation.\footnote{We acknowledge that the test period does not cover a complete annual seasonal cycle, which limits the assessment of forecasting performance across all seasons.}

    The forecasting target was restricted to property-damage claims, and the predictor selection was performed before model tuning, using the $ 2022$–$ 2024$ training period exclusively. The total claim count was excluded because it contains the property-damage count itself. Several portfolio-composition variables were also strongly related to portfolio size. During training, the share of policies held by legal entities correlated with the number of active policies at $r=-0.96$, the share of policies older than four years at $r=-0.92$, and the mean vehicle age at $r=0.69$. Based on this screening, four predictors were retained for subsequent evaluation, namely, the current property-damage claim count $C_t$, the number of active policies $P_t$, the weekday of the target day $D_{t+1}$, and the public-holiday indicator $H_{t+1}$. 

    Table~\ref{tab:predictors} summarizes the training-period evidence for the four retained predictors. Pearson correlation $r$ measures linear association, while mutual information (MI) can also detect nonlinear dependence. To assess whether an observed MI value was larger than expected by chance, the target was randomly permuted $500$ times and MI was recomputed for each permutation. The reported permutation $p$-value is the proportion of shuffled samples with MI at least as large as the observed value. We also use $\Delta R^2$ to measure the additional variance explained when a predictor is added to an existing linear model. Weekday is the strongest calendar predictor, accounting for approximately $24\%$ of the target variance. Public holidays provide additional information beyond weekday alone, increasing explained variance by $6.9$ percentage points. The current claim count $C_t$ has only weak linear correlation with the next-day count ($r=0.10$), but its MI is significant relative to the permutation baseline, indicating nonlinear dependence. The active-policy count $P_t$ serves as the measure of portfolio exposure and contributes an additional $4.7$ percentage points of explained variance beyond the calendar predictors. The four retained predictors also contain little overlapping linear information. All variance inflation factors are below $2$, indicating low multicollinearity.

    \begin{table}[t]
    \centering
    \caption{Training-period evidence for the selected predictors ($2022$--$2024$).}
    \label{tab:predictors}
    \footnotesize
    \setlength{\tabcolsep}{4pt}
    \renewcommand{\arraystretch}{1.15}
        \begin{tabular}{@{}lccp{3.7cm}@{}}
        \hline
        Input & Pearson $r$ & MI$^{*}$ ($p$) & Main evidence \\
        \midrule
        $D_{t+1}$ & -- & $0.117$ ($0.002$) &
        24\% of target variance explained by weekday \\
        
        $H_{t+1}$ & $-0.25$ & $0.040$ ($0.002$) &
        $\Delta R^2=0.069$ beyond weekday \\
        
        $C_t$ & $0.10$ & $0.048$ ($0.022$) &
        Nonlinear dependence despite weak $r$ \\
        
        $P_t$ & $0.21$ & $0.023$ ($0.15$) &
        $\Delta R^2=0.047$ beyond calendar \\
        
        \hline
        \multicolumn{4}{@{}l@{}}{%
        \parbox{\dimexpr\columnwidth-2\tabcolsep\relax}{%
        \scriptsize
        \textit{Note:} $^{*}$MI is measured in nats; $p$ is obtained from $500$ target permutations. $\Delta R^2$ denotes the additional explained variance after adding the predictor. Pearson $r$ is omitted for weekday because its numerical labels represent categorical values.}} \\
        \hline
        \end{tabular}
    \end{table}
        The claim series also contains temporal structure, particularly at short and weekly time scales. Much of this temporal structure is already represented by the calendar variables, the current claim count, and the gradual growth in portfolio exposure. We therefore do not add an explicit set of previous claim counts as input features. Instead, information from previous days is represented through the internal memory of the recurrent and reservoir models.

\subsection{Forecasting Task}
\label{subsec:problem}
    The resulting task is one-day-ahead forecasting of the property-damage claim count. For forecast origin $t$, the target is $y_t=C_{t+1}$, and the feature vector is $\mathbf{x}_t=(C_t,P_t,D_{t+1},H_{t+1})$. Here, $D_{t+1}\in\{1,\dots,7\}$ and $H_{t+1}\in\{0,1\}$ refer to the target day, which is known when the forecast is issued. Thus, all information used by the models is available at forecast time. Model fitting and cross-validation use the training period, the validation period is used for configuration selection, and the test period is reserved for final evaluation. During the test period, newly observed claims may enter subsequent forecasts as $C_t$, as they would in operational forecasting, but no model parameters are updated. 

\begin{figure}[t]
    \centering
    \includegraphics[width=0.85\columnwidth]{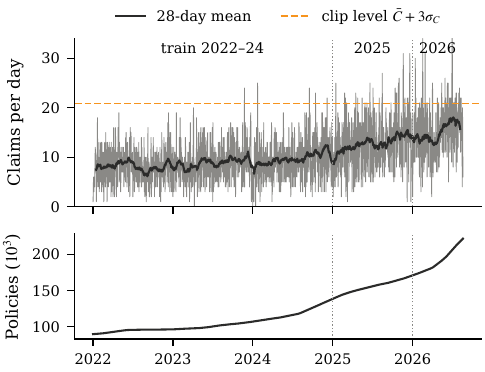}
    \caption{Daily claim counts (gray), their 28-day moving mean (black), and the number of active policies. The dashed horizontal line marks the training-derived clipping threshold used in the quantum input encoding, defined in~\eqref{eq:enc}. Dotted vertical lines separate the training, validation (2025), and test (2026) periods.}
    \label{fig:data}
    \end{figure}

    Fig.~\ref{fig:data} shows an important change in scale across these periods. The number of active policies increased from about $90{,}000$ at the beginning of the study to $222{,}000$ by August $2026$, while mean daily claims rose from $8.86$ in the training period to $14.89$ in the test period. Claim frequency per policy nevertheless remained comparatively stable, between $0.076$ and $0.085$ claims per thousand policy-days, indicating that much of the increase in the claim level reflects portfolio growth. This exposure shift is important for interpreting the later $2026$ results because the models are evaluated at portfolio sizes substantially above those observed during training. The orange dashed horizontal line marks the training-derived clipping threshold of the quantum input encoding in~\eqref{eq:enc}.\footnote{As the claim level rises in $2026$, an increasing fraction of observations reaches this training-derived clipping threshold (see Table~\ref{tab:encoding} in Subsec.~\ref{subsec:obj1}).}

        To confirm the final predictor set before the nonlinear models were evaluated, we compared three nested feature sets, given as $F_2=\{C_t,D_{t+1}\}$, $F_3=F_2\cup\{P_t\}$, and $F_4=F_3\cup\{H_{t+1}\}$. The comparison employed ridge regression and Poisson and NB2 GLMs, trained on the $2022$--$2024$ training period and evaluated on the $2025$ validation period. Forecast accuracy was measured primarily by the RMSE, expressed in claims per day, with lower values indicating more accurate predictions. Adding portfolio exposure $P_t$ reduced the $2025$ RMSE from $6.30$--$6.32$ to $4.94$--$5.01$ claims/day across all three models. Adding the holiday indicator $H_{t+1}$ produced a further reduction of $0.06$--$0.14$ claims/day, giving RMSEs of $4.87$--$4.88$. The four-variable set $F_4$ was therefore fixed for the subsequent analysis. Subsequent independent hyperparameter searches for the ESN, RNN, GRU, and XGBoost models, each considering $F_2$--$F_4$, consistently identified $F_4$ as the best-performing feature set.

\subsection{Objectives and Evaluation Metrics}\label{subsec:metrics}
    The study addresses two complementary objectives. First, we evaluate whether QRC can forecast next-day insurance claims competitively with classical models under the same chronological protocol. Second, we examine whether hardware-aware choices of reservoir topology and physical-qubit layout preserve the predictions of a frozen QRC after transfer from ideal simulation to a noisy quantum processor, and compare this preservation across three layout strategies. These objectives require distinguishing forecast accuracy from transfer distortion.

    Let $e_t=\hat y_t-y_t$ denote the forecast error at time step $t$, where $\hat y_t$ is the predicted claim count and $y_t$ is the observed count. For an evaluation period containing $n$ observations, indexed by $t=1,\ldots,n$, our primary measure of forecast accuracy is the root-mean-square error (RMSE), defined as

    \begin{align}
        \RMSE=\sqrt{\frac{1}{n}\sum_{t=1}^{n}e_t^2},\label{eq:rmse}
    \end{align}
    expressed in claims per day. We additionally report the mean absolute error (MAE), $\MAE=\frac{1}{n}\sum_{t=1}^{n}|e_t|$, and the mean forecast bias 
    \begin{align}
        \bar e=\frac{1}{n}\sum_{t=1}^{n}e_t,
    \end{align}
    where $\bar e>0$ indicates systematic overprediction and $\bar e<0$ underprediction. 
    
    To distinguish systematic bias from errors in day-to-day variation, we also report the centered RMSE (cRMSE),
    \begin{align}
       \cRMSE=\sqrt{\frac{1}{n}\sum_{t=1}^{n}(e_t-\bar e)^2},\label{eq:crmse}
    \end{align}
    which satisfies the identity $\RMSE^2=\bar e^{\,2}+\cRMSE^2$. Thus, two models can follow daily fluctuations similarly while differing in overall RMSE because one is systematically shifted upward or downward. We refer to this systematic shift as the forecast level. For pairwise comparisons evaluated on the same observations, we use the paired RMSE difference
    \begin{align}
        \Delta_{A,B}=\RMSE_A-\RMSE_B,
    \end{align}
    so that $\Delta_{A,B}<0$ favors model $A$. Uncertainty in pairwise RMSE differences is assessed using a paired circular-block bootstrap with $10{,}000$ resamples~\cite{Politis1992Circular}. The same resampled date blocks are used for both models, preserving short-range temporal dependence. We use $14$-day blocks for the primary analysis and $7$- and $28$-day blocks as sensitivity checks. The $95\%$ confidence intervals are calculated from the bootstrap distribution of $\Delta_{A,B}$. An interval excluding zero provides evidence of a difference in forecasting RMSE between the compared models.   

    To measure how faithfully the quantum processor reproduces the predictions of the corresponding ideal reservoir, we define the transfer distortion. For a frozen QRC, let $\delta_t=\hat y_t^{Q}-\hat y_t^{I}$ denote the difference between the hardware prediction $\hat y_t^{Q}$ and the ideal prediction $\hat y_t^{I}$ at time step $t$. The \emph{transfer distortion} is defined as
    \begin{align}
        d=\sqrt{\frac{1}{n}\sum_{t=1}^{n}\delta_t^2},
    \label{eq:transfer_distortion}
    \end{align}
    expressed in claims per day.
    
    A smaller value of $d$ indicates that hardware execution more closely reproduces the predictions of the ideal forecaster. Since the classical readout is trained using ideal simulated features and remains unchanged during hardware execution, $d$ quantifies the prediction-level distortion introduced by transferring the frozen forecaster to the QPU. Because the readout weights transform the measured reservoir features into predictions, $d$ does not directly quantify the loss of quantum information within the reservoir.    
    A natural alternative is to retrain the readout on the hardware features themselves. In a diagnostic comparison on the hardware runs, we trained the same ridge readout once on QPU features and once on ideal features of the same days and compared their forecasts. The median transfer distortion between them was 1.4 claims/day instead of 3.9 for the frozen readout, but the QPU-trained readout remained less accurate than the ideal-trained one in 60 of 63 runs, with median RMSE increases of $0.10$--$0.23$ claims/day across candidates. Retraining thus compensates much of the rescaling of the features but not the information lost on the processor. We therefore keep the readout frozen and use $d$ to compare layouts.

   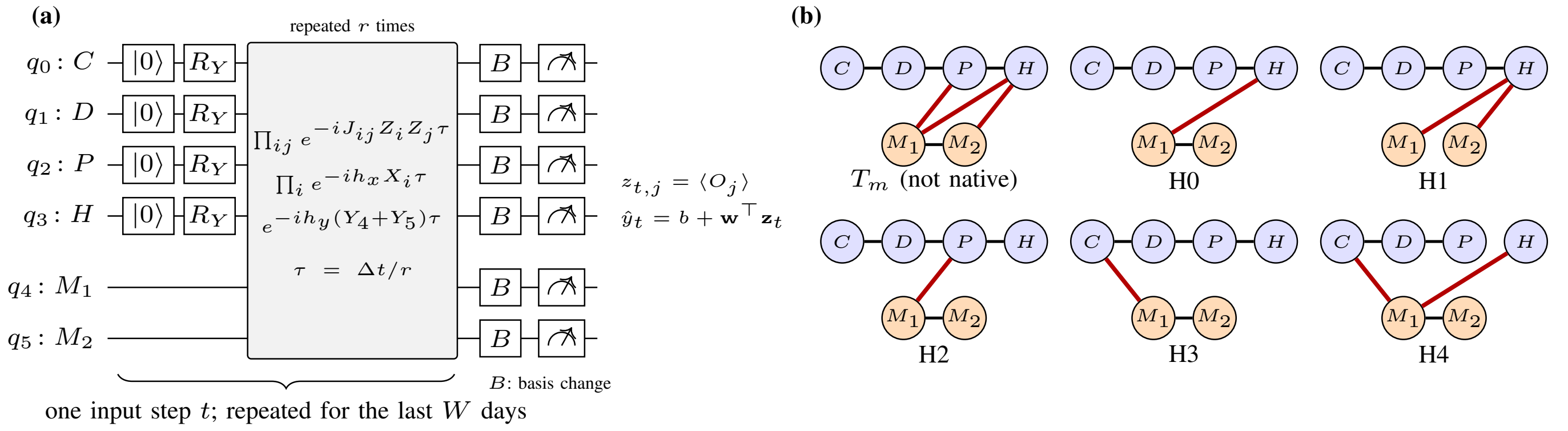
\begin{figure*}[!t]
    \centering
    \resizebox{\textwidth}{!}{%
    \begin{tikzpicture}[font=\scriptsize,
      gate/.style={draw, fill=white, minimum height=3.6mm, minimum width=5mm, inner sep=1pt},
      big/.style={draw, fill=black!5, rounded corners=1pt, align=center, inner sep=1.5pt}]
    \node[font=\scriptsize\bfseries] at (-0.6,3.25) {(a)};
    \foreach \y/\n in {2.8/$q_0\!:C$,2.3/$q_1\!:D$,1.8/$q_2\!:P$,1.3/$q_3\!:H$,0.6/$q_4\!:M_1$,0.1/$q_5\!:M_2$} {
      \node[anchor=east] at (0,\y) {\n};
      \draw (0,\y) -- (4.8,\y);
    }
    \foreach \y in {2.8,2.3,1.8,1.3} {
      \node[gate] at (0.4,\y) {$|0\rangle$};
      \node[gate] at (1.0,\y) {$R_Y$};
    }
    \node[big, font=\tiny, minimum height=31mm, text width=19.5mm] at (2.4,1.45)
      {$\prod_{ij}e^{-iJ_{ij}Z_iZ_j\tau}$\\[4pt] $\prod_i e^{-ih_xX_i\tau}$\\[4pt] $e^{-ih_y(Y_4+Y_5)\tau}$\\[6pt] $\tau=\Delta t/r$};
    \node[font=\tiny] at (2.4,3.15) {repeated $r$ times};
    \draw[decorate, decoration={brace, amplitude=4pt, mirror}] (0.1,-0.25) -- (3.4,-0.25)
      node[midway, below=4pt, align=center] {one input step $t$; repeated for the last $W$ days};
    \foreach \y in {2.8,2.3,1.8,1.3,0.6,0.1} { \node[gate, minimum width=4mm] at (3.85,\y) {$B$}; }
    \foreach \y in {2.8,2.3,1.8,1.3,0.6,0.1} {
      \node[draw, fill=white, minimum height=3.6mm, minimum width=4.5mm] at (4.45,\y) {};
      \draw (4.32,\y-0.07) arc (180:0:0.13);
      \draw[->] (4.45,\y-0.07) -- (4.55,\y+0.11);
    }
    \node[align=left, anchor=west, font=\tiny] at (3.6,-0.35) {$B$: basis change };
    \node[font=\tiny, align=left, anchor=west] at (4.9,1.45) {$z_{t,j}=\langle O_j\rangle$\\[2pt] $\hat y_t=b+\mathbf w^\top\mathbf z_t$};
    \begin{scope}[xshift=7.2cm, yshift=0.15cm,
      inj/.style={circle,draw,fill=blue!12,inner sep=0pt,minimum size=4.2mm,font=\tiny},
      mem/.style={circle,draw,fill=orange!28,inner sep=0pt,minimum size=4.2mm,font=\tiny},
      e/.style={thick}, b/.style={very thick, red!70!black}]
    \node[font=\scriptsize\bfseries] at (-0.35,3.1) {(b)};
    \newcommand{\nodesat}[2]{%
      \begin{scope}[shift={(#1,#2)}]
      \node[inj] (0) at (0,0.75) {$C$}; \node[inj] (1) at (0.6,0.75) {$D$};
      \node[inj] (2) at (1.2,0.75) {$P$}; \node[inj] (3) at (1.8,0.75) {$H$};
      \node[mem] (4) at (0.6,0) {$M_1$}; \node[mem] (5) at (1.2,0) {$M_2$};
      \end{scope}}
    \nodesat{0}{1.85}
    \draw[e] (0)--(1) (1)--(2) (2)--(3) (4)--(5); \draw[b] (2)--(4) (3)--(4) (3)--(5);
    \node at (0.9,1.5) {$T_m$ (not native)};
    \nodesat{2.45}{1.85}
    \draw[e] (0)--(1) (1)--(2) (2)--(3) (4)--(5); \draw[b] (3)--(4);
    \node at (3.35,1.5) {H0};
    \nodesat{4.9}{1.85}
    \draw[e] (0)--(1) (1)--(2) (2)--(3); \draw[b] (3)--(4) (3)--(5);
    \node at (5.8,1.5) {H1};
    \nodesat{0}{0.15}
    \draw[e] (0)--(1) (1)--(2) (2)--(3) (4)--(5); \draw[b] (2)--(4);
    \node at (0.9,-0.2) {H2};
    \nodesat{2.45}{0.15}
    \draw[e] (0)--(1) (1)--(2) (2)--(3) (4)--(5); \draw[b] (0)--(4);
    \node at (3.35,-0.2) {H3};
    \nodesat{4.9}{0.15}
    \draw[e] (0)--(1) (1)--(2) (4)--(5); \draw[b] (0)--(4) (3)--(4);
    \node at (5.8,-0.2) {H4};
    \end{scope}
    \end{tikzpicture}
    }
    \caption{(a) One replay-window step: injection qubits are reset and encoded with $R_Y(\theta_{t,j})$, all qubits evolve under $r$ Trotter slices of (\ref{eq:trotter}), and after $W$ steps are measured in one of two settings. (b) Logical coupling graphs; red edges connect the injection and memory registers. $T_m$ maximizes memory in simulation but contains triangular connections that are not supported by the heavy-hex topology; H0--H4 are five-edge trees that embed without SWAP gates.}
        \label{fig:reservoir}
    \end{figure*}
    
    While transfer distortion quantifies the overall deviation from ideal predictions, it does not distinguish between systematic prediction shifts and observation-dependent deviations. To characterize these components, we first define the \emph{mean hardware-induced shift} as
    \begin{align}
    \label{eq:hi_shift}
        \bar\delta=\frac{1}{n}\sum_{t=1}^{n}\delta_t,
    \end{align}
    which indicates whether hardware predictions are systematically shifted upward or downward relative to the ideal QRC. The squared transfer distortion can be decomposed as $d^2=\bar\delta^{\,2}+\operatorname{Var}(\delta_t)$, where $\operatorname{Var}(\delta_t)=n^{-1}\sum_{t=1}^{n}(\delta_t-\bar\delta)^2$ represents the variance of observation-dependent deviations. To quantify the contribution of the systematic shift, we define the \emph{level-shift share} as
    \begin{align}
        \rho_{\rm lvl}=\frac{\bar\delta^{\,2}}{d^2},\qquad d>0.
    \end{align}
    Values close to one indicate that hardware execution primarily introduces a systematic shift while largely preserving the day-to-day prediction pattern, whereas values close to zero indicate that observation-dependent deviations account for most of the squared transfer distortion.
    
    Forecast accuracy and transfer distortion are therefore evaluated separately: RMSE, MAE, bias, and cRMSE quantify agreement with observed claims, while $d$, $\bar\delta$, and $\rho_{\rm lvl}$ characterize how faithfully the frozen forecaster reproduces its ideal predictions on hardware.

\section{Quantum Reservoir Model}
\label{sec:qrm}
    The QRC implemented in this work uses six qubits, shown in Fig.~\ref{fig:reservoir}a. Qubits $q_0,\dots,q_3$ form the injection register $I$ and encode the four selected inputs $C_t$, $D_{t+1}$, $P_t$ and $H_{t+1}$, whereas $q_4$ and $q_5$ form the memory register $M$. At each input step, the injection qubits are reset and re-prepared, whereas the memory qubits retain the state propagated from the preceding step. This separation between an input register and an evolving quantum reservoir follows the general QRC framework of Fujii and Nakajima \cite{Fujii2017QRC}. The input variables are encoded through single-qubit $R_Y$ rotations, with rotation angles $\theta_{t,j}, j=0,\dots,3$, determined by the corresponding feature values. The resulting injection state is given as
    \begin{align}
        \rho_I(\bm\theta_t)=\bigotimes_{j=0}^{3}R_Y(\theta_{t,j})|0\rangle\langle0|R_Y(\theta_{t,j})^\dagger .
    \end{align}
    where $\bm\theta_t$ is the resulting input-angle vector. Let $z_C$ and $z_P$ denote the current claim count and active-policy count standardized using the training-period mean and standard deviation. These variables, together with weekday and holiday information, are mapped to rotation angles as
    \begin{align}
         \!\bm\theta_t\!=\!\left(\!a\,\frac{\mathrm{clip}(z_C,-3,3)}{3},\frac{2\pi(D_{t+1}\!-\!1)}{7},a\,\frac{z_P}{3\!+\!|z_P|},\pi H_{t+1}\!\right).
    \label{eq:enc}
    \end{align}
    The weekday is encoded cyclically over a full rotation, while the binary holiday indicator is mapped to either $0$ or $\pi$. For the continuous variables, the standardized claim count is clipped to $[-3,3]$, whereas portfolio exposure is encoded using a scaled softsign transformation \cite{Bergstra2009Softsign}, which smoothly compresses large values while preserving their ordering. This transformation avoids the saturation caused by direct clipping, which would otherwise map large portfolio values in $2025$ and $2026$ to identical rotation angles. The scale parameter $a$, selected during reservoir optimization, controls the magnitude of the claim-count and exposure rotations. The increasing frequency of claim-count clipping in $2026$ (Fig.~\ref{fig:data}) is further examined in Subsec.~\ref{subsec:obj1} when interpreting the test-period behavior of QRC.      
    
    After input preparation, all six qubits evolve under the fixed transverse-field Ising Hamiltonian with an additional memory-local $Y$ field
    \begin{align}
        H=\sum_{(i,j)\in E}J_{ij}Z_iZ_j+h_x\sum_{i=0}^{5}X_i+h_y(Y_4+Y_5),
    \label{eq:ham}
    \end{align}
    where $E$ denotes the logical coupling graph, $J_{ij}$ is the interaction strength between connected qubits $i$ and $j$, and $X_i$, $Y_i$, and $Z_i$ denote the corresponding Pauli operators acting on qubit $i$. The global transverse-field strength is given by $h_x$, and $h_y$ is the additional field applied to the two memory qubits.  
    The $ZZ$ interactions create correlations between connected qubits, allowing information from the injection register to reach the memory qubits. The transverse $X$ field changes the quantum states during evolution, while the additional $Y$ field acts only on the two memory qubits, providing another contribution to their dynamics. This allows us to vary the memory dynamics without modifying the input encoding or the interaction graph.    
    The strength $h_y$ is treated as a reservoir hyperparameter, with $h_y=0$ also considered during optimization. The Hamiltonian evolution is approximated using a first-order Trotter product formula \cite{Trotter1959Product}, implemented as
    \begin{align}
        \!U\!=\!\!\left[\prod_{(i,j)\in E}\!\!e^{-iJ_{ij}Z_iZ_j\tau}\prod_i\!e^{-ih_xX_i\tau}\prod_{m=4,5}\!\!e^{-ih_yY_m\tau}\!\right]^r ,
        \label{eq:trotter}
    \end{align}
    where $\Delta t$ denotes the dimensionless evolution time per input step, $r$ is the number of Trotter slices, and $\tau=\Delta t/r$ is the duration of each slice. The operator products follow the same fixed order in simulation and on hardware. Increasing $r$ improves the approximation of the Hamiltonian evolution but also increases circuit depth and the number of quantum gates. We therefore optimize $r$ as a reservoir design parameter.

    Because quantum measurement disturbs the reservoir state, RWP is used to replay the recent input history before each measurement. For a forecast at time $t$, the memory register is initialized to $|00\rangle$, and the most recent $W$ inputs, $\bm\theta_{t-W+1},\dots,\bm\theta_t$, are replayed sequentially before measurement.\footnote{Two retained candidates were originally designed for continuous operation. For hardware deployment, they are reproduced by RWP windows longer than their measured stable washout time; see Subsec.~\ref{subsec:candidates}.} The replay-window length $W$ determines how much recent history is supplied to the reservoir and is treated as a model hyperparameter, selected during reservoir optimization.    
    After each input step, the injection qubits are reset, while the memory qubits retain their evolved state for the next step. The density-matrix form is given by
    \begin{align}
        \rho_t=U\left[\rho_I(\bm\theta_t)\otimes\rho_M(t-1)\right]U^\dagger ,
    \end{align}
    with $\rho_M(t)=\mathrm{Tr}_I(\rho_t)$. 
    After the final replay step, the reservoir is measured to extract its output features. Since the recent input history is replayed for each measurement, the memory state does not need to be preserved between measurements. The reservoir features are given by the expectation values of the selected observables, $z_{t,j}=\mathrm{Tr}(O_j\rho_t)$, where $O_j$ denotes the $j$th observable and $\rho_t$ is the reservoir state after the last replay step.    

    The final forecast is produced by a classical ridge readout \cite{Hoerl1970Ridge},
    \begin{align}
        \hat y_t=b+\mathbf w^\top\mathbf z_t,
    \label{eq:readout}
    \end{align}
    where $\mathbf z_t$ is the vector of measured reservoir features (observables), $\mathbf w$ contains their weights, and $b$ is the intercept. The readout is fitted on the features of the training period. Each feature is standardized with its training mean $\bar{\mathbf z}$ and standard deviation $\mathbf s$\footnote{Features with zero variance in the training period would be excluded from the standardization and assigned zero weight; no such feature occurred.}, giving the matrix $Z=(X-\mathbf 1\bar{\mathbf z}^\top)\,\mathrm{diag}(\mathbf s)^{-1}$, and the target vector is centered with its training mean $\bar y$. The penalized coefficients are
    \begin{align}
        \bm\beta=\left(Z^\top Z+\lambda I\right)^{-1}Z^\top\left(\mathbf y-\bar y\,\mathbf 1\right),
    \label{eq:ridge}
    \end{align}
    where $\lambda$ controls the strength of regularization. They are returned to the scale of the measured observables as $w_j=\beta_j/s_j$, and the intercept, which is not regularized, is recovered as $b=\bar y-\mathbf w^\top\bar{\mathbf z}$. The readout is fitted once on noiseless training-period features and then frozen. The same $\mathbf w$ and $b$ are applied without refitting to ideal, noisy-simulation, and QPU-generated features of the validation and test periods.\footnote{The readout is not constrained to nonnegative values, but all forecasts produced in our experiments were nonnegative.}

\section{From Logical Design to Five Hardware Candidates} \label{sec:5candidates}
    The logical reservoir design was completed before any use of the test period. With the predictor set, six-qubit register and input encoding already fixed, the remaining design concerned the interaction graph, Hamiltonian parameters, temporal protocol, measured observables and hardware suitability. Model fitting and cross-validation used the training set, while the validation set was used only to rank the resulting configurations. This process corresponds to Stage~2 in Fig.~\ref{fig:workflow}.

\subsection{Hardware-Native Interaction Graphs}
    The initial memory-optimized reservoir used the seven-edge graph $T_m$ shown in Fig.~\ref{fig:reservoir}b. Its delayed-input memory, quantified using the memory-capacity metric defined in \eqref{eq:mcch} (Subsec.~\ref{subsec:selection}), was the highest among the simulated designs. However, the topology cannot be embedded directly into the heavy-hex coupling maps of the IBM processors considered, which have a maximum vertex degree of three and contain no triangular connections \cite{Chamberland2020HeavyHex}. An exhaustive search found no native embedding of $T_m$ on the IBM processors considered during development. Rather than introducing routing and additional SWAP gates, we constructed five six-qubit trees, H0--H4, that preserve the four injection and two memory qubits while differing in how information reaches the memory register. H0 forms a chain with one injection--memory connection; H1 connects the holiday input directly to both memory qubits; H2 connects the exposure input to an interacting memory pair; H3 moves the claims input to the memory boundary; and H4 uses one memory qubit as a hub. 
    
    All five graphs embed natively on the heavy-hex processor and therefore allow the reservoir dynamics to be implemented without routing. The number of possible native placements nevertheless differs substantially between graphs. On IBM Kingston, H0 and H3 admit $1072$ native embeddings, H1 and H4 admit $516$, and H2 admits $264$. The logical topology therefore affects not only the reservoir dynamics but also the number of physical-qubit layouts available during hardware deployment.

\subsection{Reservoir Parameter and Observable Optimization}
    Because the meaning of a coupling depends on the graph on which it acts, the interaction strengths optimized for $T_m$ were not transferred directly to H0--H4. Each graph was instead given the same search budget over coupling strengths $J_{ij}$, followed by local refinement. This initial topology screen showed that H2, H3 and H4 retained substantially more of the memory capability of $T_m$ than H0 and H1, indicating that both the presence and placement of injection--memory connections influence temporal information storage.
    The reservoir dynamics were then re-optimized over the interaction strengths and fields in \eqref{eq:ham}, the Trotter depth $r$, and the ridge penalty $\lambda$. The search included $r\in\{1,2,3, 4\}$, several transverse-field strengths $h_x$, and both zero and nonzero memory-field values $h_y$. The additional $Y$ field was therefore not imposed on the model but retained only when supported by the search. Most optimized branches preferred $r=3$, although shallower reservoirs remained competitive and were retained when they reduced circuit cost.
    Each candidate used one of the predefined observable sets. These ranged from the eight single-qubit $X$ and $Z$ observables on the four injection qubits to larger sets that additionally included the memory correlator $Y_4X_5$ or all single-qubit $X$, $Y$ and $Z$ observables. The observable set was selected separately for each candidate. Observables were retained only when their training-data variation exceeded the expected finite-shot measurement uncertainty, as quantified by \eqref{eq:observable_shot_noise} and the resolvability criterion defined in Subsec.~\ref{subsec:selection}. The complete hyperparameter search spaces, retained parameter settings, intermediate numerical results, and supporting scripts are provided in the Data and Code Availability materials.

    \begin{table*}[t]
    \centering
    \caption{Frozen Configurations and Hamiltonian Parameters of the Five Candidates}
    \label{tab:cand}
    \setlength{\tabcolsep}{3pt}
    \renewcommand{\arraystretch}{1.1}
    
    \resizebox{\textwidth}{!}{%
        \begin{tabular}{@{}llcccccccccccc|ccccccc@{}}
        \toprule
        & & \multicolumn{12}{c}{Frozen configuration}
        & \multicolumn{7}{c}{Ising couplings} \\
        \cmidrule(lr){3-14}
        \cmidrule(l){15-21}
        & Graph & $W$ & $r$ & $a$ & $\Delta t$
        & $h_x$ & $h_y$ & $\lambda$ & $m$
        & CV & Val. & CZ & $t_{\rm c}$ ($\mu$s)
        & $J_{01}$ & $J_{04}$ & $J_{12}$ & $J_{23}$
        & $J_{24}$ & $J_{34}$ & $J_{45}$ \\
        \midrule
        C1 & H3 & $4$ & $3$ & $0.25$ & $1.6$
        & $0.7775$ & $-0.2935$ & $10^{-4}$ & $8$
        & $3.46$ & $4.71$ & $240$ & $16$
        & $1.5178$ & $0.8391$ & $0.2776$ & $0.2399$
        & -- & -- & $-0.4389$ \\
        
        C2 & H3 & $2$ & $3$ & $0.25$ & $1.6$
        & $0.7775$ & $-0.2935$ & $10^{-1}$ & $7$
        & $3.54$ & $4.94$ & $120$ & $8$
        & $1.5178$ & $0.8391$ & $0.2776$ & $0.2399$
        & -- & -- & $-0.4389$ \\
        
        C3 & H2 & $2$ & $2$ & $0.75$ & $1.6$
        & $0.4907$ & $+0.3975$ & $10^{2}$ & $7$
        & $3.62$ & $5.06$ & $80$ & $8$
        & $-0.0834$ & -- & $0.2473$ & $1.9287$
        & $0.5131$ & -- & $-0.3865$ \\
        
        C4 & H0 & $117^{*}$ & $4$ & $0.75$ & $1.6$
        & $0.5801$ & $-0.6943$ & $10^{2}$ & $9$
        & $3.67$ & $5.03$ & $9360$ & $499$
        & $-1.6179$ & -- & $-0.1998$ & $-1.6306$
        & -- & $-0.4896$ & $0.5392$ \\
        
        C5 & H2 & $64^{*}$ & $2$ & $0.75$ & $1.6$
        & $0.4907$ & $+0.3975$ & $10^{2}$ & $8$
        & $3.59$ & $4.98$ & $2560$ & $233$
        & $-0.0726$ & -- & $0.2473$ & $1.9287$
        & $0.5131$ & -- & $-0.3865$ \\
        
        \bottomrule
        \end{tabular}%
        }
    
    \smallskip
    
    \parbox{\textwidth}{%
    \scriptsize
    \textit{Notes:}
    $W$ denotes the replay-window length;
    $^{*}$ indicates a continuous-operation design deployed using replay;
    $r$ is the number of Trotter slices;
    $a$ is the input-encoding scale;
    $\Delta t$ is the dimensionless reservoir evolution interval;
    $h_x$ and $h_y$ are the transverse- and memory-field strengths, respectively;
    $\lambda$ is the ridge-regularization parameter;
    $m$ is the number of measured observables;
    $t_{\rm c}$ is the scheduled circuit duration (in $\mu$s) of the longer measurement setting on IBM Kingston.
    A dash indicates that the corresponding Ising coupling is absent from the logical graph.
    CV denotes the mean chronological cross-validation RMSE over the training set, and Val. denotes the validation RMSE; both are expressed in claims/day.
    CZ is the total number of CZ gates per feature vector across the two measurement settings.
    }
    \end{table*}

\subsection{Multi-Objective Selection and Noisy Screening}
\label{subsec:selection}
    The parameter searches produced separate continuous (CONT) and RWP candidate pools for each of the five logical graphs H0--H4. Because forecasting accuracy, temporal memory, measurement robustness, and hardware cost represent distinct and potentially competing objectives, the reservoirs were not ranked using a single aggregate score. Instead, four complementary selection rules were used. 
    \begin{itemize}
        \item \textit{R1 -- Forecast accuracy:}
         Reservoir configurations were evaluated by chronological cross-validation (CV) on the training set, and R1 selects the configuration with the lowest mean CV RMSE within each graph--protocol pool. 
        \item \textit{R2 -- Delayed-input memory:}
        This criterion measures how well information about earlier inputs can still be recovered from the current reservoir features and is evaluated independently of forecasting accuracy. Its formal definition is given below.
        \item \textit{R3 -- Measurement robustness:}
        Among reservoirs for which every retained observable varies across the training data by more than its expected shot noise at 1024 shots, R3 selects the configuration with the lowest mean CV RMSE on the training set.
        \item \textit{R4 -- Circuit efficiency:}
        We select the least expensive reservoir whose CV RMSE remains within one standard error of the best configuration in the same graph--protocol pool. Circuit cost is compared using, in order, SWAP count, CZ count, number of measurement settings, circuit depth, and scheduled duration.
    \end{itemize}    
    For R2, delayed-input memory was measured independently of forecasting accuracy. Following the standard linear memory-capacity framework \cite{Jaeger2002STM,Dambre2012Capacity}, the four encoded input sequences were independently permuted to remove the temporal structure of the original insurance series while preserving the marginal distribution of each sequence. The experiment was repeated using five independently permuted probe sequences, and the resulting scores were averaged. For each sequence $j$ and delay $k$, a separate linear readout was trained on the training-set reservoir features to reconstruct the input value $u_j(t-k)$ from the reservoir state at time $t$.
    Let $\hat u_{j,k}(t)$ denote this reconstructed value. The memory capacity at delay $k$ was evaluated on the $N_{\rm val}=365$ validation endpoints as
    \begin{align}
    \small
        \!\mathrm{MC}_{j,k}\!=\!\max\left[\mathrm{Corr}^{2}\!\left(\!u_j(t\!-\!k),\hat u_{j,k}(t)\right)\!-\!\frac{1}{N_{\rm val}\!-\!1},\,0\right].
    \label{eq:mc}
    \end{align}
    The subtraction corrects for the expected finite-sample squared correlation between unrelated sequences \cite{Dambre2012Capacity}. A value close to one indicates that the input from $k$ steps earlier can still be reconstructed accurately from the current reservoir features.
    
    The selection score averages this memory over the four encoded input sequences and sums it over $1$--$20$ delays 
    \begin{align}
        \mathrm{MC}_{\rm s}^{1:20}=\frac{1}{4}\sum_{j=1}^{4}\sum_{k=1}^{20}\mathrm{MC}_{j,k}.
    \label{eq:mcch}
    \end{align}

    The upper limit of $20$ steps is a fixed diagnostic horizon used to compare reservoirs on the same scale; it is not interpreted as the measured memory length of the reservoir. The original memory-optimized graph $T_m$ reached $\mathrm{MC}_{\rm s}^{1:20}=1.27$ and therefore serves as the reference point for H0--H4. 
    
    Memory capacity should not be confused with washout. 
    Unlike memory capacity, which measures how well past inputs can be reconstructed, washout measures how quickly the reservoir loses dependence on its initial state. The memory of RWP reservoirs is limited by the replay window $W$, but stable washout was required for CONT reservoirs under R2.
    
    Regarding R3, measurement robustness was assessed at two levels. First, we asked whether each observable could be resolved reliably with a finite number of shots. For observable $j$, let $\sigma_j$ denote its standard deviation across the training data, representing how much that observable naturally varies between different inputs. This variation was compared with the expected finite-shot sampling uncertainty. Because Pauli observables have outcomes $\pm1$, the standard deviation of an expectation value estimated from $N$ shots is
    \begin{align}
        \sigma_{\mathrm{shot},j}=\sqrt{\frac{1-\mu_j^2}{N}},
    \label{eq:observable_shot_noise}
    \end{align}
    where $\mu_j$ is the mean expectation value of observable $j$ over the training data and $N=1024$. If the natural variation of the observable was smaller than its expected sampling uncertainty, $\sigma_j < \sigma_{\mathrm{shot},j}$, the observable was considered insufficiently resolvable for hardware use. Second, we estimated how the remaining measurement noise propagates through the ridge readout. Since the forecast is a weighted sum of the measured observables, even a small measurement error can noticeably affect the prediction when the corresponding readout weight is large. Assuming independent shot noise across observables, we used a forecast screening proxy, given as  
    \begin{align}
        \tilde{\sigma}_f =
        \left[
            \sum_j
            \left(
                \beta_j \frac{\sigma_{\mathrm{shot},j}}{\sigma_j}
            \right)^2
        \right]^{1/2},
    \end{align}
    where $\beta_j$ is the ridge coefficient on the standardized feature scale of observable $j$. Within each of the ten graph--protocol pools defined by H0--H4 under CONT and RWP operation, $\tilde{\sigma}_{\rm f}$ was used only as a tie-breaker after the resolvability and cross-validation criteria were applied.\footnote{Observables measured in the same setting are not strictly independent; the largest observed single-shot correlation was $0.49$ between $Z_2$ and $Z_3$. Including covariance did not change the selected candidates.} This criterion was motivated by an early hardware diagnostic on the validation set. The holiday observable $X_3$ had a training-set standard deviation of only $0.017$, below its expected shot-noise level of approximately $0.031$ at $1024$ shots. Because the fitted ridge coefficient on this feature was large, including $X_3$ increased the validation RMSE to about $39$ claims/day, whereas removing it reduced the error to about $5.5$. The final observable sets therefore exclude features whose natural variation is not sufficiently resolved at the available shot count (Table~\ref{tab:qpu_settings}).

    Applying R1--R4 across the ten graph--protocol pools yielded $40$ configurations, corresponding to $33$ unique reservoirs. These were evaluated using a calibrated Aer simulator with an IBM Kingston-derived noise model accounting for gate errors, thermal relaxation, and finite-shot sampling. Each reservoir was tested at different Trotter depths and shot counts. The configurations were then compared using Pareto-based screening of validation RMSE, circuit cost, and sampling budget. The Pareto-nondominated configurations represented five distinct reservoirs, which were retained as the frozen candidates C1--C5.

\subsection{Five Frozen Candidates} \label{subsec:candidates}
    The five candidates, C1–C5, were selected to represent distinct trade-offs among forecasting accuracy, memory capacity, measurement robustness, and hardware cost, rather than the five configurations with the lowest forecasting errors (Table~\ref{tab:cand}). C1 is the accuracy-oriented H3 reservoir selected by R1 and R3. C2 uses the same logical graph but substantially reduces circuit cost under R4. C3 is a shallow H2 reservoir selected for measurement robustness. C4 is a continuous H0 design, while C5 is the memory-oriented H2 candidate selected by R2. H1 and H4 contributed no candidate because none of their operating points lay on the Pareto front of the noisy screen; the best H4 point (replay, $W=14$, RMSE $4.99$ at $840$ CZ) is dominated by an H3 point with RMSE $4.98$ at $120$ CZ.
    
    Candidates C4 and C5 were optimized under CONT operation, in which the memory register is never reset between inputs. On hardware, this is not possible: measuring the features destroys the reservoir state, so a single continuous evolution cannot supply a forecast for every day. We therefore replay a sufficiently long input history before each forecast. The required replay length follows from the washout time, which measures how quickly the reservoir forgets its initial memory state~\cite{Lukosevicius2009RC}. The reservoir is initialized in each of five different memory states ($|00\rangle$, $|11\rangle$, $|{+}{+}\rangle$, $(|00\rangle+|11\rangle)/\sqrt2$ and the maximally mixed state) and driven by the same training inputs. At every step we record the largest pairwise trace distance between the memory states,
    \begin{align}
        D_{\rm tr}(t)=\max_{a<b}\tfrac12\bigl\|\rho_M^{(a)}(t)-\rho_M^{(b)}(t)\bigr\|_1 ,
    \label{eq:washout_trace}
    \end{align}
    and the largest pairwise distance between the corresponding feature vectors,
    \begin{align}
        D_{\rm feat}(t)=\max_{a<b}\bigl\|\mathbf z^{(a)}_t-\mathbf z^{(b)}_t\bigr\|_2 .
    \label{eq:washout_feat}
    \end{align}    
    The washout length is set as the first step after which $\max\{D_{\rm tr}(t),D_{\rm feat}(t)\}<\varepsilon$ remains satisfied, with $\varepsilon=0.01$ chosen as a heuristic threshold for the design search. This gives replay windows of $W=117$ for C4 and $W=64$ for C5. On the validation period, the replayed ideal forecasts reproduce the continuous predictions with RMSE differences of $0.0001$ for C4 and $0.0024$ for C5. All five configurations were subsequently frozen before test evaluation.

\section{QPU Execution and Baselines}\label{sec:execution}

    To investigate the effect of physical-qubit placement on QRC forecasting performance and simulation-to-QPU transfer, we propose a hardware-aware layout-selection procedure and compare it with two reference policies, random-native and Qiskit placement. For each of the five frozen reservoir candidates, these policies map the same six-qubit logical circuit (Table~\ref{tab:qpu_settings}) onto physical qubits, with performance assessed through ideal simulation, calibrated noisy simulation, and QPU execution.
    Ideal reservoir features are obtained from noiseless statevector simulation. For calibrated-noise simulation, expectation values are computed under an IBM Kingston-derived noise model incorporating gate errors and thermal relaxation (relaxation acts only during gates; idle periods carry no noise). QPU features are reconstructed directly from finite-shot measurement counts. The same frozen ridge readout is applied in all cases, and no Estimator primitive or readout-error mitigation is used.\footnote{In preliminary validation runs of C1 on IBM Kingston (ideal RMSE $4.76$), EstimatorV2 gave an RMSE of $6.67$ without and $9.12$ with resilience level~1, against $6.38$ for SamplerV2 on the same layout, and M3 readout mitigation raised it to $8.57$. We therefore retained the unmitigated Sampler protocol, which also preserves the raw measurement statistics needed to inspect observable-level distortions and partial hardware snapshots.}

    \begin{itemize}
        \item \emph{Hardware-aware (HA).}
        The HA procedure selects the physical layout in several stages. First, all native embeddings of the candidate's logical graph are enumerated on the current IBM Kingston coupling map, and only layouts that compile without routing are retained. These layouts are then ranked using a calibration-aware workload proxy,
    \begin{align}
        L(\ell)\!=\!-\!\sum_{g\in G_\ell}\log(1-\epsilon_g)\!-\!\sum_{q\in M_\ell}\log(1\!-\!\epsilon^{\rm RO}_q),
    \label{eq:proxy}
    \end{align}
    where $\ell$ denotes a physical layout, $\epsilon_g$ is the reported error of gate occurrence $g$, and $\epsilon^{\rm RO}_q$ is the readout error of measured qubit $q$. Because gates are counted as often as they occur in the compiled workload, the score reflects both hardware quality and the actual circuit structure. Smaller values indicate a more favorable layout.    
    The six best layouts according to this proxy are then evaluated with the Kingston-calibrated noise model on $20$ training endpoints using $64$ shots per measurement setting. For each layout, we compute the transfer distortion $d$ from \eqref{eq:transfer_distortion} between its noisy predictions and the corresponding ideal predictions. The two layouts with the smallest $d$ are subsequently tested on the QPU using $100$ evenly spaced training endpoints and $256$ shots per setting. The layout with the smaller measured value of $d$ is frozen for the final evaluation. All layout-selection stages use only training-period inputs and targets.
    \item \emph{Random native.}
        All native embeddings that avoid qubits flagged as faulty are shuffled, and the first layout that compiles without routing is selected. Calibration error rates are not used in this policy.
    \item \emph{Qiskit.}
        A candidate circuit is transpiled at optimization level~1~\cite{JavadiAbhari2024Qiskit}, and the initial layout selected by the transpiler is extracted and then fixed for routing-free compilation. 
        VF2Layout was used for C1, C2, C3, and C5, whereas C4 used the trivial layout because its chain topology could be embedded directly onto the first six physical qubits.

\end{itemize}

    \begin{table}[t]
    \centering
    \caption{Hardware Execution Settings of the Five Frozen Candidates
    ($2048$ Shots per Setting)}
    \label{tab:qpu_settings}
    \setlength{\tabcolsep}{3pt}
    \renewcommand{\arraystretch}{1}
    
    \resizebox{\columnwidth}{!}{%
        \begin{tabular}{@{}lcllc@{}}
        \toprule
         & Rounds & Measurement settings & Observables
         & Qiskit layout \\
        \midrule
        C1 & 5 &
        \texttt{XXXZYX}, \texttt{ZZZZZZ} &
        $X_{0\text{--}2}$, $Z_{0\text{--}3}$, $Y_4X_5$ &
        VF2Layout \\
        
        C2 & 5 &
        \texttt{XXXZ}, \texttt{ZZZZ} &
        $X_{0\text{--}2}$, $Z_{0\text{--}3}$ &
        VF2Layout \\
        
        C3 & 5 &
        \texttt{XXXZ}, \texttt{ZZZZ} &
        $X_{0\text{--}2}$, $Z_{0\text{--}3}$ &
        VF2Layout \\
        
        C4 & 3 &
        \texttt{XXXXYX}, \texttt{ZZZZZZ} &
        $X_{0\text{--}3}$, $Z_{0\text{--}3}$, $Y_4X_5$ &
        Trivial \\
        
        C5 & 3 &
        \texttt{XXXX}, \texttt{ZZZZ} &
        $X_{0\text{--}3}$, $Z_{0\text{--}3}$ &
        VF2Layout \\
        \midrule
        
        \multicolumn{5}{@{}p{\dimexpr\columnwidth-2\tabcolsep\relax}@{}}{%
        \scriptsize
        \textit{Notes:} Measurement settings specify the measurement bases of $q_0$--$q_5$; qubits not listed are not measured. Qubits $q_0$--$q_3$ encode $C$, $D$, $P$, and $H$, respectively, while $q_4$ and $q_5$ constitute the memory register. The $X_3$ readout is omitted from C1--C3 based on the R3 candidate-selection criterion. Each measurement setting uses 2048 shots.
        } \\
        \bottomrule
        \end{tabular}%
    }
    \end{table}   
    Each complete evaluation contains 232 test dates and two measurement settings, giving 464 circuits per layout. For each frozen candidate, the three layout policies are evaluated in paired rounds, with their jobs submitted in the same round. The HA jobs are submitted after the QPU probe (median $21$~min after the other two layout policies), and execution times can additionally differ because of queueing and provider scheduling. IBM calibration may also change between layout selection and execution; these changes are recorded but cannot be controlled experimentally. The calibration was updated between layout selection and execution in $40$ of the $63$ evaluations. The physical layout of each evaluation, its split into classical pre-processing, QPU and post-processing time, and whether the calibration changed before execution are reported in Table~\ref{tab:runtiming} in the Appendix.

    \begin{table*}[t]
    \centering
    \setlength{\tabcolsep}{15pt}
    \renewcommand{\arraystretch}{0.9}
    \caption{Forecast Accuracy on the Test Period (232 Days, Claims/Day)}
    \label{tab:forecast}
        \begin{tabular*}{\textwidth}{@{}@{\extracolsep{\fill}}lcccccc@{}}
        \toprule
        Model & RMSE & MAE & Bias & cRMSE &
        $\Delta$ vs ridge & $\Delta$ vs XGBoost (sq.) \\
        \midrule
        ridge
        & $5.36$ & $4.12$ & $+0.33$ & $5.35$
        & ---
        & $-0.92$ [$-1.63$, $-0.15$] \\
        
        NB2 GLM
        & $5.57$ & $4.28$ & $+0.86$ & $5.50$
        & $+0.21$ [$-0.03$, $+0.46$]
        & $-0.71$ [$-1.54$, $+0.25$] \\
        
        ESN$^\dagger$
        & $5.60$ & $4.31$ & $+0.41$ & $5.40$
        & $+0.24$ [$+0.12$, $+0.35$]
        & $-0.68$ [$-1.32$, $+0.02$] \\
        
        GRU$^\dagger$
        & $5.76$ & $4.45$ & $-1.50$ & $5.54$
        & $+0.40$ [$-0.03$, $+0.80$]
        & $-0.52$ [$-0.89$, $-0.10$] \\
        
        vanilla RNN$^\dagger$
        & $6.11$ & $4.71$ & $-2.15$ & $5.56$
        & $+0.75$ [$+0.22$,$+1.24$]
        & $-0.17$ [$-0.41$, $+0.09$] \\
        
        XGBoost (squared)
        & $6.28$ & $4.83$ & $-3.67$ & $5.10$
        & $+0.92$ [$+0.16$, $+1.61$]
        & --- \\
        
        Poisson GLM
        & $6.47$ & $5.06$ & $+2.93$ & $5.76$
        & $+1.10$ [$+0.43$, $+1.85$]
        & $+0.19$ [$-0.95$, $+1.56$] \\
        
        XGBoost (Poisson)
        & $6.63$ & $5.14$ & $-4.29$ & $5.05$
        & $+1.27$ [$+0.44$, $+2.04$]
        & $+0.35$ [$+0.27$, $+0.42$] \\
        
        \midrule
        
        C1, ideal
        & 5.76 & 4.36 & $-$2.80 & 5.03
        & +0.39 [$-$0.30, +1.02]
        & $-$0.53 [$-$0.68, $-$0.36] \\
        
        C2, ideal
        & $6.12$ & $4.64$ & $-3.12$ & $5.27$
        & $+0.76$ [$+0.04$, $+1.43$]
        & $-0.16$ [$-0.28$, $-0.04$] \\
        
        C3, ideal
        & $5.94$ & $4.47$ & $-2.62$ & $5.33$
        & $+0.58$ [$-0.08$, $+1.15$]
        & $-0.34$ [$-0.51$, $-0.17$] \\
        
        C4, ideal
        & $5.66$ & $4.34$ & $-1.54$ & $5.45$
        & $+0.30$ [$-0.17$, $+0.69$]
        & $-0.62$ [$-0.97$, $-0.26$] \\
        
        C5, ideal
        & $5.75$ & $4.32$ & $-2.16$ & $5.33$
        & $+0.39$ [$-0.17$, $+0.91$]
        & $-0.53$ [$-0.76$, $-0.29$] \\
        
        \midrule
        \multicolumn{7}{@{}p{\dimexpr\textwidth-2\tabcolsep\relax}@{}}{%
        \scriptsize
        $\Delta$: RMSE of the model minus RMSE of the reference, with
        $95\%$ paired circular-block bootstrap interval using $14$-day blocks.
        $^\dagger$Mean over ten fixed seeds.
        } \\
        \bottomrule
        \end{tabular*}
    \end{table*}
    
    Because the aim of this study is to evaluate the proposed HA layout selection against random-native placement and the default Qiskit placement, rather than to optimize compilation, optimization level~1 is used for all three layout policies. An offline check with a saved IBM Kingston calibration showed that levels~2 and~3 produced the same CZ count as level~1 and circuit durations within $2\%$. The Qiskit-selected layout also remained unchanged for all candidates except C4, for which VF2Layout replaced the trivial layout. The higher optimization levels therefore provided insufficient improvement in circuit cost to justify additional hardware evaluations.
    
    As baselines, we used classical models with the same $F_4$ inputs and chronological split as the QRC, with weekday represented by its sine and cosine components. The comparison includes ridge regression; Poisson and NB2 GLMs with alternative treatments of portfolio exposure; classical reservoir and recurrent models (ESN, vanilla RNN, and GRU); and XGBoost with squared-error and Poisson objectives. Seasonal-naive, persistence, and training-mean forecasts provide additional simple references.         
    All model families follow the same three-stage chronological evaluation protocol. First, five-fold expanding-window cross-validation is performed on the training set, and the three configurations with the lowest mean CV RMSE are selected as candidates for validation. For the GLMs, these configurations represent predefined treatments of portfolio exposure rather than conventional hyperparameter tuning. Second, each of the three candidates is fitted on the full training period and evaluated on the validation set. The candidate achieving the lowest validation RMSE is selected as the final configuration for its model family. The three naive forecasts are compared directly on the validation set. Third, only the selected configuration from each model family is evaluated on the test set, without further model selection or refitting. For the ESN, RNN, and GRU, all reported validation and test scores are averaged over the same ten fixed random seeds. Table~\ref{tab:classcfg} in the Appendix reports the three retained configurations per model family, their CV and validation RMSEs, the configuration selected for test evaluation, and the corresponding number of trainable parameters.

\section{Results}\label{sec:results}

\subsection{Objective 1: Forecasting Performance}\label{subsec:obj1}    
\begin{figure}[t]
    \centering
    \includegraphics[width=0.9\columnwidth]{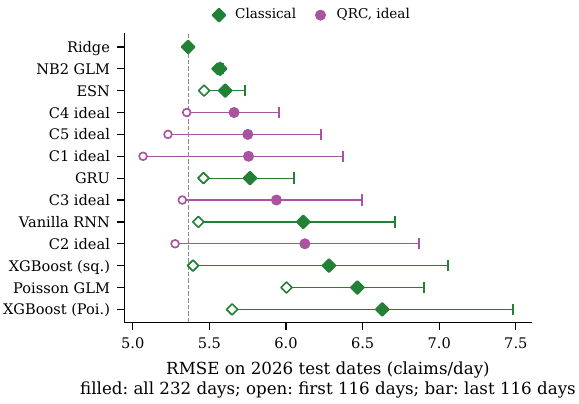}
    \caption{Test RMSE of classical models and ideal QRC candidates. Filled markers indicate the full 232-day period; open markers and bars denote the first and last 116 days. The dashed line indicates ridge RMSE.}
    \label{fig:forecast}
    \end{figure}

    Table~\ref{tab:forecast} and Fig.~\ref{fig:forecast} compare the frozen classical models and five ideal QRC candidates over the $232$-day test period. Ridge achieves the lowest RMSE of $5.36$ claims/day, while the ideal QRC candidates achieve RMSEs between $5.66$ and $6.12$ claims/day. Under the $14$-day circular-block bootstrap, the RMSE differences relative to ridge are not statistically significant at the $95\%$ confidence level for C1, C3, C4, and C5. Only C2 shows a confidence interval excluding zero, with an RMSE difference of $0.76$ claims/day. Compared with squared-error XGBoost, all five ideal QRC candidates achieve lower RMSEs, with reductions of $0.16$--$0.62$ claims/day and confidence intervals excluding zero. Their performance is also broadly comparable to that of the classical recurrent and reservoir models. The ESN achieves a slightly lower RMSE than all five QRC candidates, whereas the QRC results fall between those of the GRU and vanilla RNN.
    
    The difference in forecasting performance between ridge and the QRC candidates is largely associated with systematic prediction bias rather than centered prediction errors. C1 achieves a cRMSE of $5.03$, while the remaining QRC candidates range from $5.27$ to $5.45$, compared with $5.35$ for ridge. However, all five reservoirs exhibit negative biases ranging from $-1.54$ to $-3.12$ claims/day, whereas ridge has a bias of only $+0.33$. This difference becomes clearer when the test period is divided in half. During the first $116$ days, every ideal reservoir matches or improves on the ridge RMSE, with values of $5.07$--$5.35$ compared with $5.36$. During the second $116$ days, when the claim level is higher, the QRC RMSE rises to $5.95$--$6.87$, whereas ridge remains at $5.37$. Squared-error XGBoost shows a similar but stronger negative level bias ($-3.67$), suggesting a similar difficulty in extrapolating to claim levels beyond those observed during training.
   
    Because the QRC error increases in the second half of the test period, we next examine whether the encoded inputs also move away from the range seen during training (Table~\ref{tab:encoding}). As defined in \eqref{eq:enc}, the claims input is clipped at three training standard deviations, corresponding to approximately $20.9$ claims. This threshold is reached on only $0.5\%$ of training days, but on $7.7\%$ of validation days and $16.4\%$ of test days. In the second half of the test period, the proportion increases to $21.6\%$. Portfolio exposure shows an even clearer shift. The normalized exposure term $\theta_P/a$ ranges from $-0.30$ to $0.49$ during training but reaches $0.65$--$0.77$ in the test set. At the same time, claim frequency remains fairly stable at $0.076$--$0.085$ claims per thousand policy-days. The QRC RMSE therefore rises with portfolio growth and the resulting shift of the encoded inputs, not with a change in claim frequency. Ridge, whose linear inputs are unbounded, keeps the same RMSE in both halves of the test set ($5.36$ and $5.37$). This association does not by itself establish that the input shift causes the higher QRC RMSE; how far the forecasts move for inputs outside the training range also depends on the reservoir dynamics of each candidate.

    \begin{table}[t]
    \centering
    \caption{Inputs Relative to the Training Range}
    \label{tab:encoding}
    \setlength{\tabcolsep}{2pt}    
    \resizebox{\columnwidth}{!}{%
        \begin{tabular}{@{}lcccccc@{}}
        \toprule
        Period & $n$ & Mean $C$ & Clipped (\%) &
        Policies ($10^3$) & $\theta_P/a$ & C per $10^3$ pol. \\
        \midrule
        training
        & $1095$ & $8.86$ & $0.5$ & $89.5$--$138.2$ & $-0.30$--$0.49$ & $0.085$ \\
        
        validation
        & $365$ & $12.71$ & $7.7$ & $138.3$--$170.7$ & $0.49$--$0.65$ & $0.082$ \\
        
        test 1. half
        & $116$ & $13.51$ & $11.2$ & $170.8$--$186.8$ & $0.65$--$0.70$ & $0.076$ \\
        
        test 2. half
        & $116$ & $16.37$ & $21.6$ & $187.0$--$221.8$ & $0.70$--$0.77$ & $0.081$ \\
        \midrule
        \end{tabular}%
    }
    \vspace{1pt}
    \parbox{\columnwidth}{\scriptsize
    Clipped denotes the percentage of days on which the claim-count encoding angle reaches the training-derived clipping threshold. $\theta_P/a$ denotes the exposure encoding angle divided by the encoding scale.}
    \vspace{2pt}
    \noindent\rule{\columnwidth}{\heavyrulewidth}
    \end{table}

\subsection{Objective 2: Simulation-to-QPU Transfer} \label{subsec:obj2}
    We next evaluate how closely the ideal QRC predictions are preserved when the same reservoirs and frozen readouts are executed on IBM Kingston. The hardware experiment comprised $21$ paired rounds conducted over five days, with five rounds each for C1--C3 and three each for C4 and C5. Each round evaluated the same candidate under all three layout policies, yielding $63$ complete hardware evaluations. All runs followed the execution protocol described in Sec.~\ref{sec:execution}. Detailed per-run results, physical layouts, calibration changes and execution times are reported in Tables~\ref{tab:allruns} and~\ref{tab:runtiming} in the Appendix.
    
    \begin{table}[t]
    \centering
    \caption{Hardware Results by Candidate and Layout Policy}
    \label{tab:hw}
    \setlength{\tabcolsep}{2.5pt}
    
    \resizebox{\columnwidth}{!}{%
        \begin{tabular}{@{}llcccccc@{}}
        \toprule
        Candidate & Policy & Rounds & RMSE & Bias &
        Median $d$ & Max. $d$ & Median $\rho_{\rm lvl}$ \\
        \midrule
        \multirow{3}{*}{C1 (ideal $5.76$)}
        & HA     & $5$ & $6.36$ $\pm$ $0.55$ & $+1.72$ & $6.11$ & $7.59$ & $0.88$ \\
        & Random & $5$ & $7.77$ $\pm$ $3.81$ & $+1.00$ & $2.82$ & $16.27$ & $0.72$ \\
        & Qiskit & $5$ & $8.54$ $\pm$ $1.44$ & $+5.50$ & $9.24$ & $11.84$ & $0.83$ \\
        \midrule
        \multirow{3}{*}{C2 (ideal $6.12$)}
        & HA     & $5$ & $5.81$ $\pm$ $0.28$ & $-1.61$ & $2.55$ & $3.49$ & $0.40$ \\
        & Random & $5$ & $15.04$ $\pm$ $12.43$ & $+11.28$ & $7.25$ & $35.28$ & $0.76$ \\
        & Qiskit & $5$ & $6.55$ $\pm$ $0.32$ & $+1.11$ & $4.62$ & $6.28$ & $0.79$ \\
        \midrule
        \multirow{3}{*}{C3 (ideal $5.94$)}
        & HA     & $5$ & $5.75$ $\pm$ $0.23$ & $-1.37$ & $1.33$ & $3.23$ & $0.59$ \\
        & Random & $5$ & $5.82$ $\pm$ $0.23$ & $-0.18$ & $3.61$ & $4.09$ & $0.70$ \\
        & Qiskit & $5$ & $6.56$ $\pm$ $0.71$ & $-2.99$ & $1.70$ & $3.17$ & $0.14$ \\
        \midrule
        \multirow{3}{*}{C4 (ideal $5.66$)}
        & HA     & $3$ & $5.83$ $\pm$ $0.39$ & $+0.84$ & $3.07$ & $3.95$ & $0.72$ \\
        & Random & $3$ & $5.81$ $\pm$ $0.54$ & $+1.37$ & $2.05$ & $4.97$ & $0.93$ \\
        & Qiskit & $3$ & $6.68$ $\pm$ $0.48$ & $+3.67$ & $5.26$ & $6.12$ & $0.98$ \\
        \midrule
        \multirow{3}{*}{C5 (ideal $5.75$)}
        & HA     & $3$ & $5.72$ $\pm$ $0.17$ & $+0.69$ & $2.72$ & $4.49$ & $0.67$ \\
        & Random & $3$ & $8.11$ $\pm$ $3.82$ & $+3.24$ & $3.15$ & $13.03$ & $0.74$ \\
        & Qiskit & $3$ & $5.97$ $\pm$ $0.34$ & $+1.82$ & $4.10$ & $5.25$ & $0.80$ \\
        \midrule
        \end{tabular}%
    }    
    \vspace{1pt}
    \parbox{\columnwidth}{\scriptsize
    \textit{Notes:} Ideal test RMSE is shown in parentheses. Hardware RMSE and bias are reported as mean$\pm$SD across rounds. Median $\rho_{\rm lvl}$ denotes the median level-shift share across rounds.}    
    \vspace{2pt}
    \noindent\rule{\columnwidth}{\heavyrulewidth}
    \end{table}

    \begin{figure*}[t]
    \centering
    \includegraphics[width=0.9\textwidth]{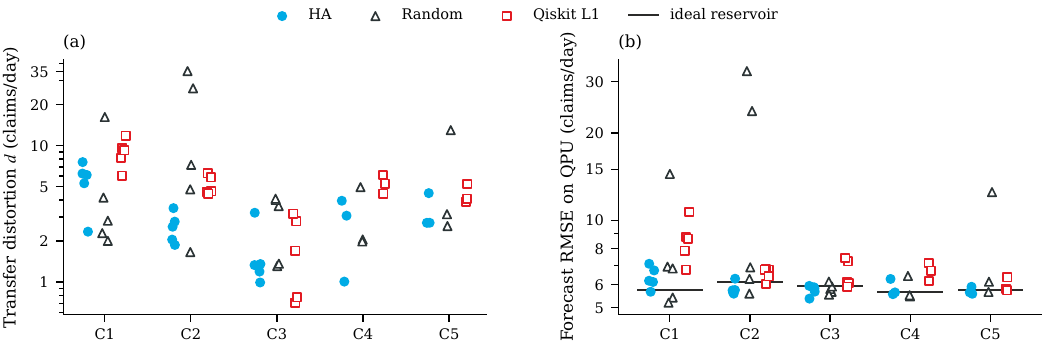}
    \caption{(a) Transfer distortion $d$ between QPU and ideal predictions. (b) Forecast RMSE on the QPU; horizontal lines mark ideal RMSE. Logarithmic axes.}
    \label{fig:hw}
    \end{figure*}
    
    \begin{figure*}[t]
    \centering
    \includegraphics[width=0.9\textwidth]{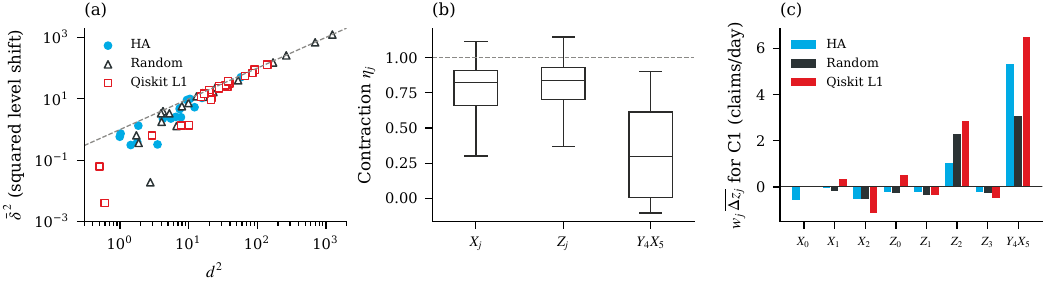}
    \caption{(a) Squared mean hardware-induced shift $\bar\delta^2$ against squared transfer distortion $d^2$ for all hardware runs; the dashed line corresponds to $\rho_{\rm lvl}=1$, where the distortion is entirely due to the mean shift. (b) Observable contraction $\eta_j$ from $z^Q_{t,j}=\eta_j z^I_{t,j}+\zeta_j$; $\eta_j=1$ indicates that the day-to-day variation of the ideal observable is preserved. (c) Contribution $w_j\overline{\Delta z_j}$ of each observable to the mean hardware-induced shift $\bar\delta$ \eqref{eq:hi_shift} of C1, averaged over rounds.}
    \label{fig:mech}
    \end{figure*}

    \begin{figure*}[h!]
    \centering
    \includegraphics[width=0.9\textwidth]{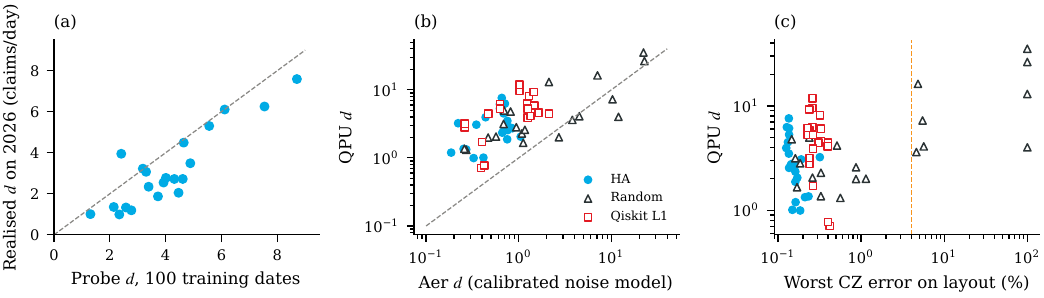}
    \caption{(a) Distortion measured by the HA probe on $100$ training dates against the
    realized test distortion. (b) Distortion predicted by the calibrated Aer noise model against the
    measured QPU distortion. (c) Realized distortion against the worst CZ error in the selected physical layout, i.e., the largest reported error among the CZ gates of the logical graph (100\% marks a coupler reported as non-functional); the dashed line marks a $4\%$ CZ error.}
    \label{fig:pred}
    \end{figure*}

    Table~\ref{tab:hw} summarizes the results by candidate and layout policy; Fig.~\ref{fig:hw} shows every individual run. The hardware behavior varies substantially across candidates and layout strategies: HA achieves lower RMSE than the Qiskit-selected layouts in all paired rounds and lower transfer distortion in $18/21$ of them. VF2Layout selects the layout with the lowest gate and readout error rates, without regard to coherence times or circuit duration. HA uses a similar error-based score only to shortlist six layouts; the noisy simulation, which includes thermal relaxation, and the QPU probe then decide, and the layout with the best error score was selected in only $3$ of the $21$ rounds. As a result, the Qiskit layouts have about half the readout error of the HA layouts (median $0.006$ against $0.011$), whereas the HA layouts use qubits with longer worst-case coherence (median minimum $T_2$ of $74$ against $40$~$\mu$s) and lower CZ errors ($0.0012$ against $0.0022$). That HA nevertheless performs better shows that ranking by error rates alone does not ensure the best transfer, and that no single calibration metric characterizes layout quality.

    The comparison with random-native placement is different. HA has the lower RMSE in $12/21$ paired rounds, and the median paired difference is only $0.27$ claims/day. A well-placed random layout can therefore match or beat HA. However, random-native placement exhibits substantially greater variability and more severe worst-case degradation. Random placement produces the four largest transfer distortions in the experiment ($d=13.0$--$35.3$ claims/day) and a worst forecast RMSE of $32.6$, whereas HA never exceeds $d=7.6$ or an RMSE of $7.1$. The main advantage of HA is therefore not a large reduction in typical error but protection against severe layout failures.\footnote{These paired differences should be read with calibration drift in mind (Table~\ref{tab:runtiming}), which affected $17$ HA, $11$ random-native, and $12$ Qiskit runs.}

    This distinction between forecast accuracy and transfer distortion is also visible at the candidate level. For C2 and C3, the mean HA hardware RMSE ($5.81$ and $5.75$) is lower than the corresponding ideal RMSE ($6.12$ and $5.94$), and the hardware RMSE stays below the ideal value in $4/5$ rounds of each candidate, even though the QPU does not reproduce the ideal predictions exactly. Both ideal reservoirs underpredict the test-set claims (bias $-3.12$ and $-2.62$, Table~\ref{tab:forecast}), and the upward shift introduced by the QPU reduces this bias to $-1.61$ and $-1.37$ (Table~\ref{tab:hw}), so two errors partly cancel. The lower hardware RMSE should therefore not be interpreted as better transfer, which is why $d$ remains the primary measure of simulation-to-QPU transfer. Among the five candidates, C3 provides the strongest overall hardware result, combining the lowest median transfer distortion of $d=1.33$ claims/day with a mean HA RMSE of $5.75$ claims/day, close to the lowest value of $5.72$ (C5).
    
    Across the hardware runs, a large part of the transfer distortion appears as a systematic shift in the predictions rather than as changes in their day-to-day pattern. In 44/63 hardware runs, $\rho_{\rm lvl}>0.5$, indicating that the squared systematic level shift accounts for more than half of the total squared transfer distortion between the QPU and ideal forecasts (Fig.~\ref{fig:mech}a). The median level-shift share over all runs is $0.74$. To examine where this shift originates, we compare each measured QPU observable with its ideal value using $z^Q_{t,j}=\eta_j z^I_{t,j}+\zeta_j $, where $\eta_j$ measures how much of the ideal variation is preserved and $\zeta_j$ represents an offset. Perfect preservation would give $\eta_j=1$ and $\zeta_j=0$. Across all runs, the median $\eta_j$ is $0.82$ for $X_j$ observables and $0.84$ for $Z_j$ observables (Fig.~\ref{fig:mech}b). The two-qubit memory correlator $Y_4X_5$ is much more strongly affected, with a median $\eta_j$ of only $0.30$.

    Because the QRC readout is linear, these observable-level errors propagate directly into the forecast. Defining $\Delta z_{t,j}=z^Q_{t,j}-z^I_{t,j}$, we have $\delta_t=\sum_j w_j\Delta z_{t,j}$, so the mean hardware-induced shift in \eqref{eq:hi_shift} decomposes exactly as $\bar\delta=\sum_j w_j\overline{\Delta z_j}$. For C1, the memory correlator $Y_4X_5$ alone contributes about $+5.3$ claims/day to the mean HA shift (Fig.~\ref{fig:mech}c), which helps explain why C1 shows the largest transfer distortion among the HA candidates. This contribution is large not because the forecast relies on $Y_4X_5$: on the standardized scale its coefficient is only $\beta_j=1.24$ claims/day per standard deviation, $16\%$ of the largest $|\beta_j|$ of C1. Rather, $Y_4X_5$ varies little in simulation (SD $0.05$), so its raw weight is large ($w_j=22.9$), and on hardware its mean changes systematically by $0.23$, about five such standard deviations, far more than finite-shot sampling can explain.    
    The size of this change follows from the observable-level fit $z^Q_{t,j}=\eta_j z^I_{t,j}+\zeta_j$, which gives $\overline{\Delta z_j}=(\eta_j-1)\,\overline{z^I_j}+\zeta_j$: contraction ($\eta_j<1$) moves the mean of any observable whose ideal mean lies away from zero, and the offset $\zeta_j$ adds to it. For $Y_4X_5$ in C1, the processor retains less than half of the ideal variation with $\eta_j=0.45$ and adds an offset of $\zeta_j=0.13$ (Fig.~\ref{fig:mech}b). 
    
    The same pattern, a large raw weight on an observable that varies little in simulation, also dominates C4, which has the second-largest median HA distortion ($d=3.07$ claims/day). There the main term comes from the injection observable $Z_2$ rather than from the memory register: with an ideal mean of $0.71$, an SD of only $0.018$ and a readout weight of $w_j=-24.4$, a mean change of about $-0.12$ raises the C4 forecasts by a median $2.9$ claims/day. For such low-variance observables, $\eta_j$ and $\zeta_j$ are poorly determined individually, whereas their combination $\overline{\Delta z_j}$ is stable.
    In both candidates, the hardware-induced shift is therefore concentrated in a few observables that vary little in simulation and thus carry large raw readout weights, and a moderate systematic change of such an observable on hardware is amplified into a shift of several claims/day, even when the observable matters little for the forecast itself.

    The hardware results raise a practical question: can poor transfer be identified before the test forecasts are executed? Fig.~\ref{fig:pred} compares three indicators available at deployment time.

    The most informative indicator is the short QPU probe used by the HA procedure. For each selected layout, the probe evaluates 100 training dates and measures the distortion from the corresponding ideal predictions (Sec.~\ref{sec:execution}). The probe distortion is strongly correlated with the later test distortion, with Pearson $r=0.88$ (Fig.~\ref{fig:pred}a). Its level shift is even more strongly correlated with the realized level shift ($r=0.94$). These results indicate that the training-period hardware probe provides a useful empirical indicator of subsequent test-period transfer distortion, although the observed correlations do not establish a general performance guarantee.
    
    Calibrated Aer simulation provides a weaker indication (Fig.~\ref{fig:pred}b). Although the simulated and measured distortions are related, Aer underestimates the observed distortion by a median factor of $4.2$. Part of this gap is built into the noise model, which omits readout errors and relaxation during idle periods (Sec.~\ref{sec:execution}). The gap shows directly in the contraction of the memory correlator, where Aer retains $\eta_j=0.89$ for C1 and $0.41$ for C4, against $0.45$ and $0.15$ on the processor (HA runs). Effects that no calibration snapshot captures, such as crosstalk and calibration drift, add to the remaining gap \cite{Klimov2018Fluctuations,Proctor2020Drift}.    
    Calibration data are nevertheless useful for identifying clearly poor layouts. Fig.~\ref{fig:pred}c shows that the largest distortions occur when the selected layout contains a highly degraded CZ coupler. All four severe random-layout failures used a coupler with a reported CZ error above $4\%$, three of them a coupler reported as non-functional, whereas no HA or Qiskit layout exceeded $0.4\%$. Random placement produced these failures because it avoids faulty qubits but does not rank otherwise native layouts by coupler quality.
    
    \begin{figure*}[t]
    \centering
    \includegraphics[width=\textwidth]{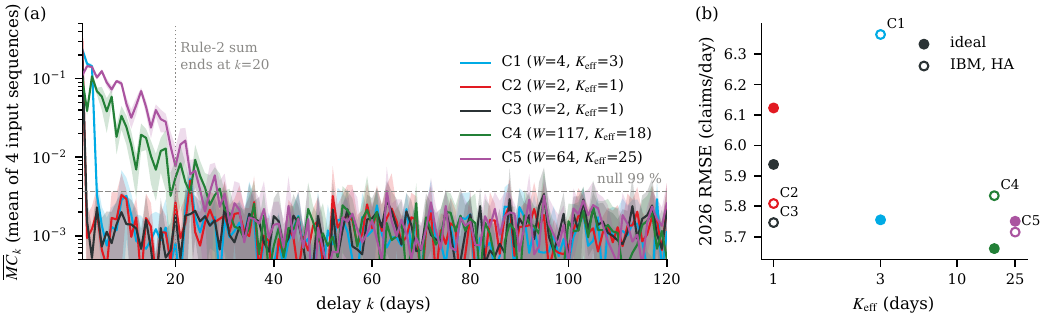}
    \caption{Delayed-input memory of the five frozen candidates.
    (a) Mean memory score $\overline{\mathrm{MC}}_k$ as a function of delay; bands show $\pm$ SD over five probe seeds. The horizontal dashed line is the $99^{\text{th}}$ percentile of the scores obtained after permuting the targets and is used as the noise threshold. The vertical dotted line marks the end of the $k=1$--$20$ range used by Rule R2.
    (b) Effective memory horizon $K_{\rm eff}$ against test RMSE for the ideal forecast (filled) and mean HA hardware result (open).}
    \label{fig:memory}
    \end{figure*}

    \begin{table}[t]
    \centering
    \caption{Memory of the Frozen Candidates and Their Test Results}
    \label{tab:memory}
    \footnotesize
    \setlength{\tabcolsep}{3pt}
    \begin{tabular}{@{}lrcrcccr@{}}
    \toprule
     & $W$ & $\mathrm{MC}_{\rm s}^{1:20}$ & $K_{\rm eff}$ & Ideal RMSE & HA RMSE & HA $d$ & Circuit $\mu$s \\
    \midrule
    C1 & $4$ & $0.56$ ($0.01$) & $3$ & $5.76$ & $6.36$ & $6.11$ & $16$ \\
    C2 & $2$ & $0.34$ ($0.01$) & $1$ & $6.12$ & $5.81$ & $2.55$ & $8$\\
    C3 & $2$ & $0.25$ ($0.01$) & $1$ & $5.94$ & $5.75$ & $1.33$ & $8$ \\
    C4 & $117$ & $0.66$ ($0.11$) & $18$ & $5.66$ & $5.83$ & $3.07$ & $499$ \\
    C5 & $64$ & $1.35$ ($0.06$) & $25$ & $5.75$ & $5.72$ & $2.72$ & $233$ \\
    \midrule
    
    \multicolumn{8}{@{}p{\dimexpr\columnwidth-2\tabcolsep\relax}@{}}{%
    \scriptsize
    $\mathrm{MC}_{\rm s}^{1:20}$: Rule-R2 memory score,
    mean (SD) over five seeds.
    $K_{\rm eff}$: effective memory horizon in days based on the $99^{\text{th}}$-percentile permutation threshold; Ideal RMSE \& HA RMSE: test RMSE; HA $d$: median transfer distortion; circuit duration is the scheduled duration of the longer measurement setting.
    } \\
    \bottomrule
    \end{tabular}
    \end{table}

\subsection{Memory and Information-Processing Capacity}
\label{subsec:memory}
    Rule R2 favored reservoirs with stronger \emph{linear} delayed-input memory, which raises a natural question: did the candidates with longer memory also forecast better or transfer more reliably to hardware? Memory at each delay $k$ was measured using the $\mathrm{MC}_{j,k}$ score defined in \eqref{eq:mc}, for $k=1,\dots,120$, averaged over the four encoded input sequences and five seeds. To distinguish genuine memory from small correlations arising by chance, we also calculated the same score after randomly permuting the targets ($20$ permutations at $12$ delays between $1$ and $117$, averaged in the same way). The $99^{\text{th}}$ percentile of these permuted scores was used as a single noise threshold for each candidate. We define the effective memory horizon $K_{\rm eff}$ as the last delay of the uninterrupted run, starting at $k=1$, for which the measured memory remains above this threshold.
    
    Fig.~\ref{fig:memory}a and Table~\ref{tab:memory} show that the replay window $W$ is only an upper bound on how far into the past the reservoir can retain useful information. For the short-window candidates, the effective memory reaches the limit allowed by the window: three days for C1 and one day for C2 and C3. The long-window candidates behave differently. C4 replays $117$ days, but its measurable ideal memory extends only to about $18$ days. C5 uses a $64$-day window but retains information for about $25$ days. Beyond these points, the memory scores stay close to the level obtained from randomly permuted targets with only isolated crossings. These findings indicate that the substantially longer replay windows of C4 and C5 increase circuit depth without producing a proportional extension of the detectable ideal memory horizon.

    More memory does not translate directly into better forecasting or lower transfer distortion (Fig.~\ref{fig:memory}b). C4 and C5 have the longest effective memory and are among the strongest ideal forecasters, but C1 reaches similar ideal accuracy with only three days of memory. On hardware, C3 provides the smallest transfer distortion despite having an effective horizon of only one day, while C5 achieves good hardware accuracy despite its substantially longer replay window. Across the five candidates, neither forecasting accuracy nor transfer distortion exhibits a consistent ordering with the ideal memory horizon.
    
    How much the forecasts actually use the input history differs between candidates. In noiseless simulation, refitting each readout on single-step ($W=1$) features, which carry no input history, raises the test RMSE by $0.49$ for C1, $0.81$ for C3 and $0.96$ for C5, but only by $0.04$ for C2 and $0.01$ for C4. C4, the candidate with the longest replay window, therefore forecasts almost as well without any history. A memory ablation on IBM Kingston confirms that its history is also not preserved on hardware. With the frozen readout, the same layout and the same session, replacing the preceding $116$ inputs by scrambled ones changed the QPU RMSE only from $5.49$ to $5.53$ claims/day, and the memory correlator $Y_4X_5$ lost almost all of its ideal variation ($\eta_j\approx0.001$, against $0.82$ for $W=1$), at little cost to the forecast because its standardized weight is only $\beta_j=-0.03$. Good QPU forecasting accuracy alone is therefore not evidence that the temporal memory of the ideal reservoir has been preserved.
    
    The preceding analysis concerns linear memory only. We therefore also examined nonlinear information-processing capacity using the decomposition of Dambre \emph{et al.}~\cite{Dambre2012Capacity}. Using the same frozen reservoirs in noiseless simulation and the same train--validation protocol, we measured their ability to reconstruct linear and nonlinear functions of current and past inputs. The total capacity was similar across all five candidates, but its distribution differed: C1--C3 retained more nonlinear information about past inputs, whereas the past-input capacity of C4 and C5 was almost entirely linear.\footnote{The nonlinear past-input capacity was approximately $2.0$ for C1, about $1.0$ for C2 and C3, and only $0.08$--$0.13$ for C4 and C5.} 

\subsection{Computational Cost}\label{subsec:cost}
    The replay window has a direct effect on hardware cost. The short-window candidates C1--C3 require only $80$--$240$ CZ gates per feature vector and scheduled circuit durations of $8$--$16$~$\mu$s, whereas C4 and C5 require $9360$ and $2560$ CZ gates and durations of $499$ and $233$~$\mu$s, respectively (Table~\ref{tab:cand}). Both long-window candidates are therefore more exposed to decoherence, and the circuits of C4 exceed the shortest coherence times of its selected qubits. C4 is thus a hardware boundary case: as shown in Sec.~\ref{subsec:memory}, most of its circuit depth adds little to its forecasting accuracy.

    The QRC and the classical baselines also differ in where their cost arises. The QRC readout has only $8$--$10$ trainable parameters, but each forecast requires the execution and measurement of quantum circuits. The classical models are trained and evaluated entirely on a CPU, with model sizes ranging from $6$ coefficients for ridge and the count-regression models, through a $201$-parameter linear readout on the fixed ESN reservoir, to fully trained recurrent networks with $26$ (GRU) and $273$ (RNN) parameters (Table~\ref{tab:classcfg}).

    For a computational time comparison, the classical models and the ideal QRC simulations were re-measured after the experiment under a common single-thread CPU protocol,\footnote{Timing measurements were performed on a workstation equipped with an AMD Ryzen 5 5500U processor ($6$ cores, $12$ logical processors, $2.1$~GHz) and $15.3$~GB RAM. IBM hardware timings were taken from the job metrics and timestamps reported by the IBM Quantum cloud service; QPU time therefore excludes queue waiting and other client-side wall-clock overhead.} fitting on the training set and forecasting all $232$ test days. Ridge, the GLMs and XGBoost required $4$--$23$~ms, the ESN about $0.22$~s per seed, the vanilla RNN $2.9$~s per seed and the GRU $17.3$~s per seed. Noiseless QRC simulation required approximately $1.1$--$1.7$~s for C1--C3, $3.9$~s for C5 and $6.1$~s for C4. On IBM Kingston, the corresponding hardware evaluations required a median of $256$--$264$~s of QPU time for C1--C3, $476$~s for C5 and $725$~s for C4. The classical preparation of these evaluations takes seconds to minutes for random-native and Qiskit placement, whereas HA layout selection adds minutes of calibrated Aer screening for C1--C3 and about an hour for C4, and a QPU probe that adds $11\%$ to the QPU time of the HA evaluations (Table~\ref{tab:runtiming}). These measurements show that current QPU execution is substantially slower than both the classical baselines and ideal QRC simulation, and that both the QPU time and the classical preparation grow strongly with the replay window.

\subsection{Portability Check on IQM Emerald} \label{subsec:iqm}
    To examine whether the design procedure depends specifically on IBM's heavy-hex connectivity, we repeated the reservoir-selection pipeline for IQM Emerald, a $54$-qubit processor with a square-lattice coupling structure. The same input encoding, Hamiltonian family, selection rules, and train-validation protocol were retained, while two additional seven-edge graphs, H5 and H6, were introduced to exploit connectivity available on the IQM lattice. The resulting five rule representatives achieved CV RMSEs of $3.33$--$3.58$ and validation RMSEs of $4.72$--$4.93$. Without any further fitting, their ideal test RMSEs range from $5.48$ to $6.12$ claims/day, closely matching the range obtained by the IBM-oriented reservoirs. They also show the same systematic underprediction, with biases from $-1.26$ to $-3.01$ claims/day. This similarity is expected because the forecasting inputs and encoding are unchanged and suggests that the main forecast-level limitation is not specific to one processor topology. Because full IQM hardware access was not available for the corresponding test evaluation, the second research objective, the simulation-to-QPU transfer, cannot be assessed on Emerald in this study. The IQM analysis therefore addresses portability of the design procedure and ideal forecasting behavior only. 

\section{Discussion}\label{sec:discussion}
    A six-qubit QRC with a small frozen linear readout captured the relevant structure of the claim series and remained competitive with strong classical nonlinear baselines. Its main weakness was not a lack of predictive structure but extrapolation. As the portfolio grew, the bounded encoding of exposure moved the test inputs outside the training range, and all candidates underpredicted. On hardware, the frozen reservoirs transferred with moderate distortion when the physical layout was chosen carefully. The hardware-aware procedure was most valuable in avoiding severe layout failures, and the short QPU probe identified poor transfer before deployment. This reliability is not free, as the probe adds about $11\%$ to the QPU time of each HA evaluation, and the classical layout screening takes a few minutes for shallow reservoirs and about an hour for C4. The study suggests several rules for deploying a QRC forecaster on current hardware.
    \begin{enumerate}
    \item Study the data before designing the circuit to determine which variables justify qubit allocation (Subsec.~\ref{sec:predictors}).
    \item Choose an encoding that remains informative outside the training range. If an input can grow substantially over time, a bounded rotation may compress new values too strongly (Subsec.~\ref{subsec:obj1}).
    \item Check that each observable is measurable above the expected shot noise. Otherwise, sampling noise can dominate the feature and degrade the forecast (Subsec.~\ref{subsec:selection}).
    \item Inspect the raw readout weights. Observables that vary little in simulation receive large weights, which amplify small systematic hardware changes into large forecast shifts (Fig.~\ref{fig:mech}).
    \item Match reservoir memory to the task. Long replay windows can greatly increase circuit cost without providing proportionally more usable memory (Subsecs.~\ref{subsec:memory} and~\ref{subsec:cost}).
    \item Exclude layouts with degraded couplers before execution using current calibration data (Fig.~\ref{fig:pred}c).
    \item Probe the chosen layout directly on the QPU using training data before deployment; calibrated simulation underestimates the hardware distortion (Fig.~\ref{fig:pred}a--b).
    \item Select layouts on distortion from the ideal predictor, not on error against observations, which rewards accidental offsets (Subsec.~\ref{subsec:obj2}).
    \end{enumerate}

    The study has several limitations. It uses a single insurance portfolio, whose data alone guided the whole pipeline, so the forecasting findings should be tested on other portfolios before being generalized. The test period covers only 232 days and therefore not a full seasonal cycle. The hardware comparison consists only of 3–5 paired rounds per candidate on one processor, cannot exclude calibration drift, and uses Qiskit only at optimization level 1. All hardware results were also obtained with six-qubit reservoirs and without error mitigation, and it remains open whether the advantage of hardware-aware placement persists for larger reservoirs or mitigated execution. The role of degraded couplers in the random-layout failures and the effect of exposure encoding were identified after the experiment. Simulation-to-QPU transfer could not be tested on IQM Emerald and is left for future work.

\section{Conclusion} \label{sec:conclusion}
    The main finding of this study is that the usefulness of a quantum reservoir forecaster depends not only on the reservoir dynamics, but also on how the resulting model is transferred to physical hardware. For the insurance task considered here, the six-qubit reservoirs captured the relevant predictive structure well enough to remain competitive with established nonlinear baselines, although the data also exposed a clear limitation of the bounded input encoding when portfolio exposure moved outside the training range. Across $63$ hardware evaluations on IBM Kingston, hardware-aware layout gave lower forecast RMSE than the Qiskit level 1 mapping in all rounds and avoided the severe transfer failures observed for some random-native layouts. Most of the remaining transfer distortion appeared as a systematic forecast shift, and a short QPU probe provided a useful indication of this behavior before final deployment. The memory analysis revealed that this task does not require long temporal context. More complex temporal tasks may therefore provide a better setting in which to exploit the longer and more nonlinear memory available in some reservoir designs. The results suggest that practical QRC deployment should balance forecasting accuracy, memory requirements and circuit cost, while selecting physical layouts according to how closely their QPU predictions reproduce the corresponding ideal forecasts.
    
    For future work, the most immediate extension is to improve the encoding under portfolio growth, for example by encoding the claim frequency per policy, which remained stable over the study period, or the logarithm of the number of policies, instead of bounded transformations of the absolute quantities. Training-only QPU probes could also be used to investigate correction of systematic hardware shifts. Beyond this dataset, larger reservoirs and time series with stronger nonlinear or long-range temporal structure would provide a more demanding test of the memory and information-processing capacity of QRC.

\begin{table*}[t]
    \centering
    \caption{Classical Configurations Using the Final $F_4$ Feature Set}
    \label{tab:classcfg}
    \renewcommand{\arraystretch}{0.95}
    \setlength{\tabcolsep}{3pt}
        \begin{tabular}{@{}llccccp{10.5cm}@{}}
        \toprule
        Method & Rank & CV & Val. & Sel. & Par. & Configuration \\ 
        \midrule
        
        Baseline & -- & -- & $6.406$ & $\checkmark$ & 0 & seasonal\_naive\_7d \\
        Baseline & -- & -- & $6.666$ &  &  & training\_mean \\
        Baseline & -- & -- & $7.118$ &  &  & persistence \\
        
        ridge & $1$ & $3.671$ &$ 4.872$ & $\checkmark$ & $6$ & alpha=$10^{-4}$ \\
        ridge & $2$ & $3.671$ & $4.872$ &  &  & alpha=$10^{-3}$ \\
        ridge & $3$ & $3.671$ & $4.872$ &  &  & alpha=$10^{-2}$ \\
        
        Poisson & $1$ & $3.619$ & $4.881$ &  &  & variant=fixed\_log\_policy\_offset; gamma=$1.0$ \\
        Poisson & $2$ & $3.710$ & $4.886$ &  &  & variant=learned\_log\_policy; gamma=$0.917$ \\
        Poisson & $3$ & $3.716$ & $4.879$ & $\checkmark$ & $6$ & variant=standardized\_policy \\
        
        NB2 & $1$ & $3.624$ & $4.882$ & $\checkmark$ & $6$ & variant=fixed\_log\_policy\_offset; alpha\_dispersion=$0.0447$; gamma=$1.0$ \\
        NB2 & $2$ & $3.715$ & $4.885$ &  &  & variant=learned\_log\_policy; alpha\_dispersion=$0.0446$; gamma=$0.919$ \\
        NB2 & $3$ & $3.722$ & $4.883$ &  &  & variant=standardized\_policy; alpha\_dispersion=$0.0445$ \\
        
        ESN & $1$ & $3.642$ & $4.950 \pm 0.155$ &  &  & window\_length=$7$; reservoir\_size=$50$; requested\_connectivity=$0.1$; spectral\_radius=$0.9$; leak\_rate=$1.0$; ridge\_alpha=$0.1$ \\
        ESN & $2$ & $3.650$ & $4.848 \pm 0.058$ & $\checkmark$ & $201$ & window\_length=$7$; reservoir\_size=$200$; requested\_connectivity=$0.1$; spectral\_radius=$0.7$; leak\_rate=$1.0$; ridge\_alpha=$1.0$ \\
        ESN & $3$ & $3.650$ & $4.930 \pm 0.119$ &  &  & window\_length=$14$; reservoir\_size=$100$; requested\_connectivity=$0.5$; spectral\_radius=$0.9$; leak\_rate=$1.0$; ridge\_alpha=$1.0$ \\
        
        RNN & $1$ & $3.557$ & $5.004 \pm 0.301$ & $\checkmark$ & $273$ & window=$1$; hidden\_size=$8$; layers=$2$; learning\_r=$0.001$; weight\_decay=$0.001$; epochs=$141$ \\
        RNN & $2$ & $3.557$ & $5.035 \pm 0.328$ &  &  & window=$1$; hidden\_size=$8$; layers=$2$; learning\_r=$0.001$; weight\_decay=$1e-05$; epochs=$138$ \\
        RNN & $3$ & $3.557$ & $5.032 \pm 0.326$ &  &  & window=$1$; hidden\_size=$8$; layers=$2$; learning\_r=$0.001$; weight\_decay=$0.0001$; epochs=$138$ \\
        
        GRU & 1 & 3.451 & $5.021 \pm 0.111$ &  &  & window=$1$; hidden\_size=$1$; layers=$1$; learning\_r=$0.001$; weight\_decay=$0.0001$; epochs=$708$ \\
        GRU & $2$ & $3.529$ & $5.231 \pm 0.239$ &  &  & window=$2$; hidden\_size=$1$; layers=$1$; learning\_r=$0.001$; weight\_decay=$0.0001$; epochs=$891$ \\
        GRU & $3$ & $3.567$ & $4.946 \pm 0.105$ & $\checkmark$ & $26$ & window=$14$; hidden\_size=$1$; layers=$1$; learning\_r=$0.001$; weight\_decay=$0.0001$; epochs=$383$ \\
        
        XGB squared & $1$ & $3.420$ & $4.901$ &  &  & max\_depth=$1$; learning\_r=$0.1$; min\_child\_w=$10.0$; reg\_lambda=$10.0$; n\_estimators=$150$ \\
        XGB squared & $2$ & $3.422$ & $4.871$ & $\checkmark$ & $300$ & max\_depth=$1$; learning\_r=$0.1$; min\_child\_w=$10.0$; reg\_lambda=$1.0$; n\_estimators=$150$ \\
        XGB squared & $3$ & $3.424$ & $4.912$ &  &  & max\_depth=$1$; learning\_r=$0.03$; min\_child\_w=$10.0$; reg\_lambda=$1.0$; n\_estimators=$400$ \\
        
        XGB Poisson & $1$ & $3.458$ & $5.075$ & $\checkmark$ & $1187$ & max\_depth=$3$; learning\_r=$0.03$; min\_child\_w=$10.0$; reg\_lambda=$10.0$; n\_estimators=$150$ \\
        XGB Poisson & $2$ & $3.461$ & $5.347$ &  &  & max\_depth=$2$; learning\_r=$0.03$; min\_child\_w=$1.0$; reg\_lambda=$10.0$; n\_estimators=$150$ \\
        XGB Poisson & $3$ & $3.461$ & $5.347$ &  &  & max\_depth=$2$; learning\_r=$0.03$; min\_child\_w=$10.0$; reg\_lambda=$10.0$; n\_estimators=$150$ \\
        
        \midrule
        \multicolumn{7}{@{}p{\dimexpr\textwidth-2\tabcolsep\relax}@{}}{%
        \scriptsize
        \textit{Notes:} CV: mean RMSE over the five folds on the training set (the naive forecasts have no cross-validation step).
        Val.: RMSE on the validation set after fitting on the training set; for ESN, RNN and GRU, values are mean $\pm$ sample SD over ten seeds.
        Sel.: the configuration evaluated on the test set.
        Par.: number of trained parameters of the selected configuration: coefficients and intercept for ridge and the GLMs
        (NB2 includes the dispersion), readout weights and intercept for the ESN, whose reservoir is fixed, all network weights
        for the RNN and GRU, and leaf values for XGBoost.
        For comparison, the QRC readouts have $8$--$10$ parameters ($7$--$9$ weights and an intercept).
        Ranks $1$--$3$ are the three best of $84$ (ridge), $14$ (GLMs), $432$ (XGBoost), $20$, $160$ (ESN), $246$ (RNN) and $126$ (GRU)
        cross-validated settings.
        } \\
        \bottomrule
        \end{tabular}
    \end{table*}

\section*{Acknowledgment}
    This work was supported by the Slovak Research and Development Agency under project no.~APVV-23-0512 and the Slovak Academy of Sciences under project no.~VEGA 1/0641/26.
    
    During the preparation of this manuscript, the authors used OpenAI ChatGPT for language editing and grammar correction. The tool was not used to design the methodology or experiments, generate or analyze experimental data, or determine the scientific conclusions. All AI-assisted text was reviewed and verified by the authors, who take full responsibility for the content of the article.

\section*{Data and Code Availability}
    The code, processed data, hardware-run records, and analysis scripts required to reproduce the results of this study are publicly available at \url{https://github.com/rusnakrenata/qrc_insurance} and archived on Zenodo at \url{https://doi.org/10.5281/zenodo.23862967}.
    
\appendix[Supplementary Tables]
    Table~\ref{tab:classcfg} reports, for each classical model family, the three configurations retained after training-period cross-validation, their validation scores and the configuration selected for the final test evaluation, following the protocol of Sec.~\ref{sec:execution}. All configurations use the final $F_4$ predictor set established in Sec.~\ref{sec:predictors}. The full search spaces and random seeds are provided in the Data and Code Availability materials.
    
    Table~\ref{tab:allruns} reports all $21$ paired rounds of QPU evaluations with the corresponding ideal and calibrated Aer results.
    
    Table~\ref{tab:runtiming} lists, for each of the $63$ QPU evaluations, the physical layout, the QPU time of the evaluation and of the HA probe, whether the IBM Kingston calibration changed between layout selection and execution, and the classical pre- and post-processing time. The classical processing times were re-measured after the experiment on the workstation of Sec.~\ref{subsec:cost}, using all $12$ logical cores, by replaying the classical steps of each layout policy with the calibration snapshot of the first round of each candidate. Because these steps are identical in every round, one value is reported per candidate and layout policy. Pre covers the layout selection and the transpilation of the evaluation circuits. For HA, we measure the proxy ranking of all native embeddings, the calibrated Aer screening of the six best layouts, and the compilation and post-processing of the QPU probe. For random-native placement, we measure the time of the native embeddings and a compile check. For Qiskit, we record the time of the preset layout pass and a compile check. Post covers the reconstruction of the features from the measurement counts and the frozen readout. 

\begin{table*}[t!]
    \centering
    \caption{Ideal, Calibrated-Noise and Hardware Results of All 21 Paired Rounds (Test Dates, Claims/Day)}
    \label{tab:allruns}
    \setlength{\tabcolsep}{6.7pt}
    \renewcommand{\arraystretch}{0.95}
        \begin{tabular*}{\textwidth}{@{}@{\extracolsep{\fill}}ccc |cc|cc |cc|cc |cc|cc@{}}
        \toprule
        & & & \multicolumn{4}{c}{Hardware-aware} & \multicolumn{4}{c}{Random native} & \multicolumn{4}{c}{Qiskit} \\
        \cmidrule(lr){4-7}\cmidrule(lr){8-11}\cmidrule(lr){12-15}
        & & Ideal & \multicolumn{2}{c}{Aer} & \multicolumn{2}{c}{QPU} & \multicolumn{2}{c}{Aer} & \multicolumn{2}{c}{QPU} & \multicolumn{2}{c}{Aer} & \multicolumn{2}{c}{QPU} \\
        \cmidrule(lr){4-5}\cmidrule(lr){6-7}\cmidrule(lr){8-9}\cmidrule(lr){10-11}\cmidrule(lr){12-13}\cmidrule(lr){14-15}
        Candidate & Round & RMSE & RMSE & $d$ & RMSE & $d$ & RMSE & $d$ & RMSE & $d$ & RMSE & $d$ & RMSE & $d$ \\
        \midrule
        C1 & $1$ & $5.76$ & $5.52$ & $0.63$ & $\mathbf{6.14}$ & $6.11$ & $6.74$ & $7.07$ & $14.45$ & $16.27$ & $5.31$ & $1.26$ & $6.76$ & $\mathbf{6.04}$ \\
        C1 & $2$ & $5.76$ & $5.52$ & $0.63$ & $5.68$ & $5.31$ & $5.45$ & $0.94$ & $\mathbf{5.43}$ & $\mathbf{2.82}$ & $5.41$ & $1.02$ & $8.77$ & $9.58$ \\
        C1 & $3$ & $5.76$ & $5.51$ & $0.66$ & $7.09$ & $7.59$ & $5.39$ & $1.08$ & $\mathbf{6.93}$ & $\mathbf{2.29}$ & $5.41$ & $1.02$ & $10.68$ & $11.84$ \\
        C1 & $4$ & $5.76$ & $5.49$ & $0.71$ & $\mathbf{6.20}$ & $6.25$ & $5.18$ & $2.71$ & $6.83$ & $\mathbf{2.01}$ & $5.31$ & $1.29$ & $7.86$ & $8.14$ \\
        C1 & $5$ & $5.76$ & $5.51$ & $0.66$ & $6.72$ & $\mathbf{2.34}$ & $5.31$ & $1.26$ & $\mathbf{5.21}$ & $4.17$ & $5.30$ & $1.45$ & $8.64$ & $9.24$ \\
        \midrule
        C2 & $1$ & $6.12$ & $5.76$ & $0.81$ & $\mathbf{5.76}$ & $2.78$ & $5.66$ & $1.12$ & $6.29$ & $\mathbf{1.66}$ & $5.48$ & $1.63$ & $6.77$ & $4.62$ \\
        C2 & $2$ & $6.12$ & $5.79$ & $0.76$ & $\mathbf{5.66}$ & $\mathbf{2.55}$ & $20.37$ & $22.72$ & $23.78$ & $26.40$ & $5.63$ & $1.23$ & $6.77$ & $6.28$ \\
        C2 & $3$ & $6.12$ & $5.78$ & $0.75$ & $\mathbf{5.59}$ & $\mathbf{3.49}$ & $5.76$ & $0.82$ & $5.60$ & $4.79$ & $5.48$ & $1.63$ & $6.76$ & $4.59$ \\
        C2 & $4$ & $6.12$ & $5.67$ & $1.05$ & $\mathbf{5.74}$ & $\mathbf{2.05}$ & $8.94$ & $10.27$ & $6.89$ & $7.25$ & $5.38$ & $2.12$ & $6.04$ & $4.46$ \\
        C2 & $5$ & $6.12$ & $5.78$ & $0.75$ & $\mathbf{6.30}$ & $\mathbf{1.87}$ & $19.82$ & $22.19$ & $32.63$ & $35.28$ & $5.54$ & $1.47$ & $6.41$ & $5.86$ \\
        \midrule
        C3 & $1$ & $5.94$ & $5.81$ & $0.33$ & $5.83$ & $\mathbf{0.99}$ & $5.51$ & $3.79$ & $\mathbf{5.68}$ & $3.61$ & $5.83$ & $0.26$ & $7.23$ & $2.79$ \\
        C3 & $2$ & $5.94$ & $5.83$ & $0.27$ & $5.95$ & $1.33$ & $5.83$ & $0.27$ & $\mathbf{5.81}$ & $\mathbf{1.31}$ & $5.83$ & $0.26$ & $7.41$ & $3.17$ \\
        C3 & $3$ & $5.94$ & $5.84$ & $0.25$ & $5.68$ & $1.36$ & $10.75$ & $11.92$ & $\mathbf{5.55}$ & $4.01$ & $5.79$ & $0.39$ & $6.14$ & $\mathbf{0.71}$ \\
        C3 & $4$ & $5.94$ & $5.85$ & $0.22$ & $\mathbf{5.38}$ & $3.23$ & $5.83$ & $0.26$ & $5.92$ & $1.36$ & $5.77$ & $0.42$ & $6.09$ & $\mathbf{0.77}$ \\
        C3 & $5$ & $5.94$ & $5.86$ & $0.19$ & $\mathbf{5.89}$ & $\mathbf{1.19}$ & $5.76$ & $4.50$ & $6.15$ & $4.09$ & $5.78$ & $0.40$ & $5.91$ & $1.70$ \\
        \midrule
        C4 & $1$ & $5.66$ & $5.56$ & $0.44$ & $6.28$ & $3.95$ & $5.54$ & $0.57$ & $\mathbf{5.48}$ & $\mathbf{2.05}$ & $5.53$ & $0.63$ & $7.14$ & $6.12$ \\
        C4 & $2$ & $5.66$ & $5.58$ & $0.35$ & $\mathbf{5.65}$ & $\mathbf{3.07}$ & $5.52$ & $0.69$ & $6.44$ & $4.97$ & $5.53$ & $0.63$ & $6.70$ & $5.26$ \\
        C4 & $3$ & $5.66$ & $5.57$ & $0.42$ & $5.57$ & $\mathbf{1.01}$ & $5.57$ & $0.47$ & $\mathbf{5.52}$ & $1.99$ & $5.55$ & $0.47$ & $6.19$ & $4.44$ \\
        \midrule
        C5 & $1$ & $5.75$ & $5.53$ & $0.69$ & $5.91$ & $4.49$ & $5.53$ & $0.69$ & $\mathbf{5.67}$ & $\mathbf{3.15}$ & $5.41$ & $1.25$ & $6.36$ & $5.25$ \\
        C5 & $2$ & $5.75$ & $5.53$ & $0.70$ & $\mathbf{5.59}$ & $2.72$ & $5.42$ & $1.16$ & $6.15$ & $\mathbf{2.58}$ & $5.41$ & $1.25$ & $5.83$ & $3.89$ \\
        C5 & $3$ & $5.75$ & $5.49$ & $0.85$ & $\mathbf{5.65}$ & $\mathbf{2.72}$ & $5.70$ & $2.11$ & $12.51$ & $13.03$ & $5.40$ & $1.34$ & $5.74$ & $4.10$ \\
        \midrule
        \multicolumn{15}{@{}p{\textwidth}@{}}{%
        \scriptsize \textit{Notes:} Ideal: noiseless simulation of the frozen model. Aer: simulation of the same layout with the gate and thermal-relaxation noise model exported from the calibration snapshot of the round (relaxation acts only during gates, exact expectation values, no shot noise, readout error not modeled). QPU: IBM Kingston run; bold marks the lowest QPU RMSE and the lowest QPU $d$ of the three layout policies in each round. $d$: transfer distortion relative to the ideal prediction. All values in claims/day.} \\
        \bottomrule
        \end{tabular*}
    \vspace{8mm}
    \centering
    \caption{Physical Layouts, Execution Time, HA Probe and Calibration Changes of All 63 QPU Evaluations on IBM Kingston}
    \label{tab:runtiming}
    \scriptsize
    \setlength{\tabcolsep}{2.5pt}
        \begin{tabular}[t]{@{}lcllrrrrc@{}}
        \toprule
        C & R & Policy & Physical layout & \makecell{Pre\\(s)} & \makecell{QPU\\(s)} & \makecell{Post\\(s)} & \makecell{Probe\\QPU (s)} & \makecell{Calibration\\change} \\
        \midrule
        C1 & 1 & HA & [74,73,79,93,75,59] & 200 & 264 & 0.03 & 31 & Yes \\
        C1 & 1 & Random & [87,86,85,84,88,89] & 7 & 264 & 0.03 & -- & No \\
        C1 & 1 & Qiskit & [150,149,148,147,151,138] & 11 & 264 & 0.03 & -- & No \\
        C1 & 2 & HA & [74,73,79,93,75,59] & 200 & 264 & 0.03 & 31 & No \\
        C1 & 2 & Random & [136,143,142,141,123,124] & 7 & 264 & 0.03 & -- & No \\
        C1 & 2 & Qiskit & [45,44,43,42,46,47] & 11 & 264 & 0.03 & -- & Yes \\
        C1 & 3 & HA & [74,75,59,55,73,79] & 200 & 264 & 0.03 & 31 & Yes \\
        C1 & 3 & Random & [91,90,89,88,92,93] & 7 & 264 & 0.03 & -- & Yes \\
        C1 & 3 & Qiskit & [45,44,43,42,46,47] & 11 & 264 & 0.03 & -- & Yes \\
        C1 & 4 & HA & [75,74,73,79,59,55] & 200 & 264 & 0.03 & 31 & No \\
        C1 & 4 & Random & [103,96,83,84,102,101] & 7 & 264 & 0.03 & -- & No \\
        C1 & 4 & Qiskit & [150,149,148,147,151,138] & 11 & 264 & 0.03 & -- & No \\
        C1 & 5 & HA & [73,79,93,94,74,75] & 200 & 264 & 0.03 & 31 & Yes \\
        C1 & 5 & Random & [38,29,28,27,49,50] & 7 & 264 & 0.03 & -- & No \\
        C1 & 5 & Qiskit & [45,46,47,48,37,25] & 11 & 264 & 0.03 & -- & No \\
        C2 & 1 & HA & [79,93,94,95,73,74] & 166 & 257 & 0.03 & 30 & No \\
        C2 & 1 & Random & [105,117,125,124,104,103] & 5 & 257 & 0.03 & -- & No \\
        C2 & 1 & Qiskit & [150,149,148,147,151,138] & 7 & 257 & 0.03 & -- & No \\
        C2 & 2 & HA & [74,73,79,93,75,59] & 166 & 257 & 0.03 & 30 & Yes \\
        C2 & 2 & Random & [113,114,115,99,112,111] & 5 & 257 & 0.03 & -- & Yes \\
        C2 & 2 & Qiskit & [45,44,43,42,46,47] & 7 & 257 & 0.03 & -- & Yes \\
        C2 & 3 & HA & [74,75,59,55,73,79] & 166 & 257 & 0.03 & 30 & Yes \\
        C2 & 3 & Random & [79,73,74,75,93,94] & 5 & 257 & 0.03 & -- & No \\
        C2 & 3 & Qiskit & [150,149,148,147,151,138] & 7 & 257 & 0.03 & -- & No \\
        C2 & 4 & HA & [103,104,105,117,96,83] & 166 & 257 & 0.03 & 30 & Yes \\
        C2 & 4 & Random & [87,88,89,78,86,85] & 5 & 257 & 0.03 & -- & Yes \\
        C2 & 4 & Qiskit & [57,67,66,65,47,48] & 7 & 257 & 0.03 & -- & Yes \\
        C2 & 5 & HA & [73,79,93,94,74,75] & 166 & 257 & 0.03 & 30 & Yes \\
        C2 & 5 & Random & [112,113,114,115,111,98] & 5 & 257 & 0.03 & -- & No \\
        C2 & 5 & Qiskit & [45,46,47,48,37,25] & 7 & 257 & 0.03 & -- & No \\
        C3 & 1 & HA & [141,142,143,144,136,123] & 137 & 256 & 0.03 & 30 & No \\
        C3 & 1 & Random & [83,84,85,86,77,65] & 6 & 256 & 0.03 & -- & No \\

        \end{tabular}\hfill
        \begin{tabular}[t]{@{}lcllrrrrc@{}}
        \toprule
        C & R & Policy & Physical layout & \makecell{Pre\\(s)} & \makecell{QPU\\(s)} & \makecell{Post\\(s)} & \makecell{Probe\\QPU (s)} & \makecell{Calibration\\change} \\
        \midrule
        C3 & 1 & Qiskit & [43,44,45,37,46,47] & 6 & 256 & 0.03 & -- & No \\
        C3 & 2 & HA & [65,66,67,57,68,69] & 137 & 256 & 0.03 & 30 & Yes \\
        C3 & 2 & Random & [145,144,143,136,142,141] & 6 & 256 & 0.03 & -- & No \\
        C3 & 2 & Qiskit & [43,44,45,37,46,47] & 6 & 256 & 0.03 & -- & No \\
        C3 & 3 & HA & [41,36,21,20,22,23] & 137 & 256 & 0.03 & 30 & Yes \\
        C3 & 3 & Random & [149,148,147,137,146,145] & 6 & 256 & 0.03 & -- & No \\
        C3 & 3 & Qiskit & [67,57,47,46,48,49] & 6 & 256 & 0.03 & -- & No \\
        C3 & 4 & HA & [149,150,151,138,152,153] & 137 & 256 & 0.03 & 30 & Yes \\
        C3 & 4 & Random & [71,58,51,50,52,53] & 6 & 256 & 0.03 & -- & Yes \\
        C3 & 4 & Qiskit & [67,57,47,48,46,45] & 6 & 256 & 0.03 & -- & Yes \\
        C3 & 5 & HA & [91,92,93,94,79,73] & 137 & 256 & 0.03 & 30 & Yes \\
        C3 & 5 & Random & [87,86,85,77,84,83] & 6 & 256 & 0.03 & -- & Yes \\
        C3 & 5 & Qiskit & [25,37,45,44,46,47] & 6 & 256 & 0.03 & -- & Yes \\
        C4 & 1 & HA & [80,81,82,83,96,103] & 4341 & 725 & 0.05 & 80 & Yes \\
        C4 & 1 & Random & [55,54,53,39,33,34] & 162 & 752 & 0.04 & -- & Yes \\
        C4 & 1 & Qiskit & [0,1,2,3,4,5] & 159 & 725 & 0.03 & -- & Yes \\
        C4 & 2 & HA & [74,75,59,55,54,53] & 4341 & 725 & 0.05 & 80 & Yes \\
        C4 & 2 & Random & [71,70,69,68,67,66] & 162 & 725 & 0.04 & -- & Yes \\
        C4 & 2 & Qiskit & [0,1,2,3,4,5] & 159 & 725 & 0.03 & -- & Yes \\
        C4 & 3 & HA & [95,94,93,79,73,72] & 4341 & 725 & 0.05 & 80 & Yes \\
        C4 & 3 & Random & [24,25,26,27,17,7] & 162 & 725 & 0.04 & -- & Yes \\
        C4 & 3 & Qiskit & [0,1,2,3,4,5] & 159 & 725 & 0.03 & -- & Yes \\
        C5 & 1 & HA & [95,94,93,92,79,73] & 954 & 476 & 0.04 & 53 & Yes \\
        C5 & 1 & Random & [93,79,73,72,74,75] & 53 & 476 & 0.04 & -- & Yes \\
        C5 & 1 & Qiskit & [47,46,45,44,37,25] & 59 & 476 & 0.04 & -- & Yes \\
        C5 & 2 & HA & [91,92,93,94,79,73] & 954 & 476 & 0.04 & 53 & Yes \\
        C5 & 2 & Random & [25,26,27,28,17,7] & 53 & 476 & 0.04 & -- & Yes \\
        C5 & 2 & Qiskit & [47,46,45,44,37,25] & 59 & 476 & 0.04 & -- & Yes \\
        C5 & 3 & HA & [91,92,93,94,79,73] & 954 & 476 & 0.04 & 53 & Yes \\
        C5 & 3 & Random & [131,130,129,118,128,127] & 53 & 476 & 0.04 & -- & Yes \\
        C5 & 3 & Qiskit & [49,48,47,46,57,67] & 59 & 476 & 0.04 & -- & Yes \\
        \end{tabular}
    \noindent\rule{\textwidth}{\lightrulewidth}
    
    \smallskip
    \parbox{\textwidth}{\footnotesize \textit{Notes:} C: candidate; R: round. Pre, Post: classical pre- and post-processing of one evaluation, re-measured after the experiment; same value in every round of a candidate and layout policy. QPU: QPU time charged by IBM. Probe QPU: QPU time charged for the HA probe of the round. Queue waiting is excluded. Calibration change: whether the backend calibration was updated between the snapshot used for layout selection and the start of execution.}
    
    \vspace{2pt}
    \noindent\rule{\textwidth}{\heavyrulewidth}
    \end{table*}

\bibliographystyle{IEEEtran}
\bibliography{bibliography}

\begin{IEEEbiography}
[{\includegraphics[width=1in,height=1.25in, clip,keepaspectratio]{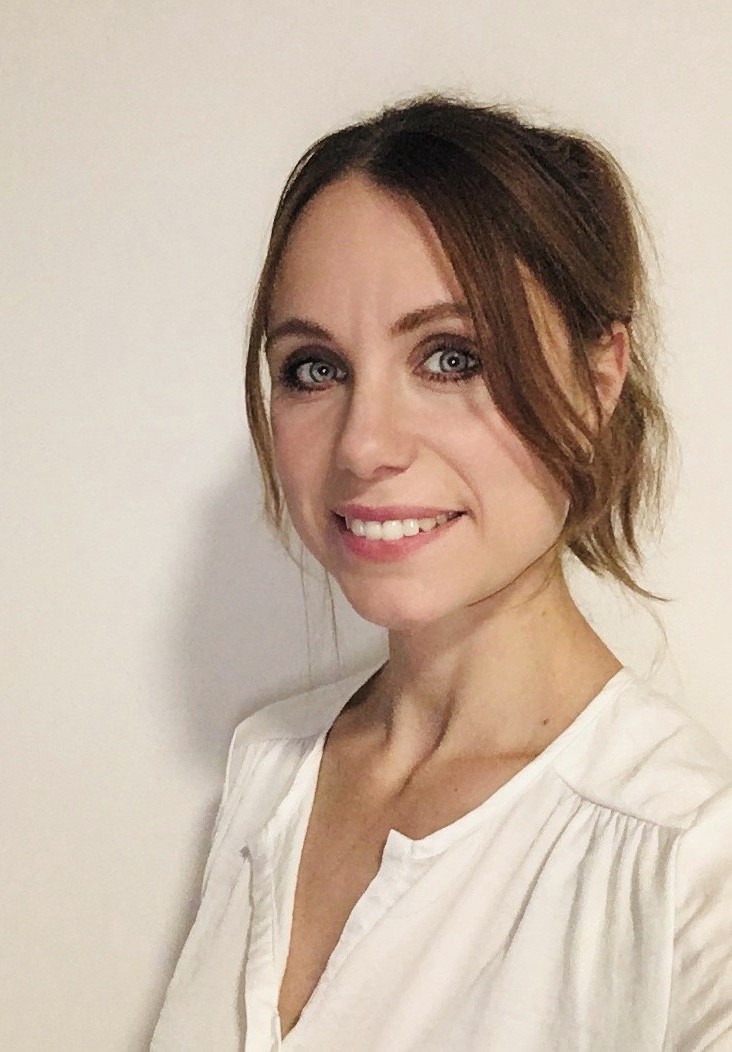}}] 
{Renáta Rusnáková} received the M.S. degree in Mathematical Methods of Information Security from Charles University, Prague, Czech Republic, in 2012. Her master's thesis focused on computational complexity and reductions among NP-hard problems.

She is currently pursuing the Ph.D. degree at the Technical University of Košice (TUKE), Slovakia, specializing in quantum computing and hybrid quantum–classical computational methods. Her research interests include quantum annealing, combinatorial optimization, quantum reservoir computing, quantum machine learning, and hardware-aware implementations of quantum algorithms, with applications in transportation optimization, cybersecurity, and time-series forecasting.

Ms. Rusnáková actively contributes to research projects and scientific publications exploring the practical capabilities of quantum and quantum-inspired computing, with particular emphasis on their application to real-world computational problems.
\end{IEEEbiography}

\begin{IEEEbiography}
[{\includegraphics[width=1in,height=1.25in,clip,keepaspectratio]{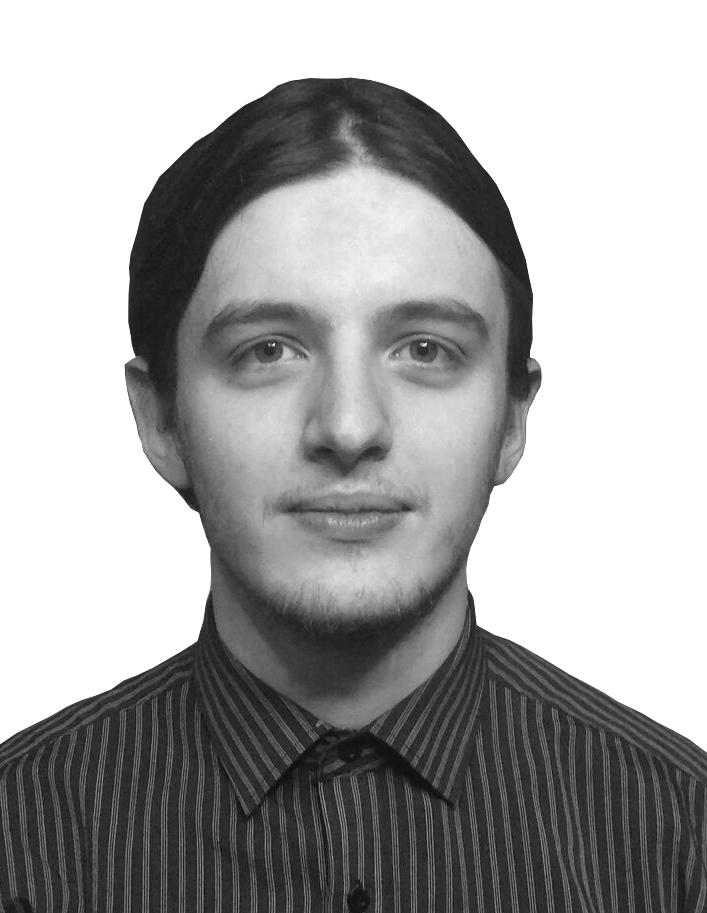}}]
{Eugen \v{S}lapak}
received the B.S. and M.S. degrees in informatics from the Technical University of Košice, Košice, Slovakia, in 2016 and 2018, respectively, and the Ph.D. degree from the same university in 2021, with a thesis on 5G HetNet physical topology design using machine learning.

Dr. \v{S}lapak is currently an Assistant Professor with the Intelligent Information Systems Laboratory, Technical University of Košice. He was a Guest Researcher at King's College London, London, U.K., under the supervision of Prof. Mischa Dohler and at the University of California, Irvine, under the supervision of Prof. Mohammad Al Faruque.  His research interests include machine learning, computer vision, radio access network simulation, and metaheuristic optimization. He serves as a regular reviewer in several recognized IEEE Transactions journals.
\end{IEEEbiography}

\begin{IEEEbiography}
[{\includegraphics[width=1in,height=1.25in,clip,keepaspectratio]{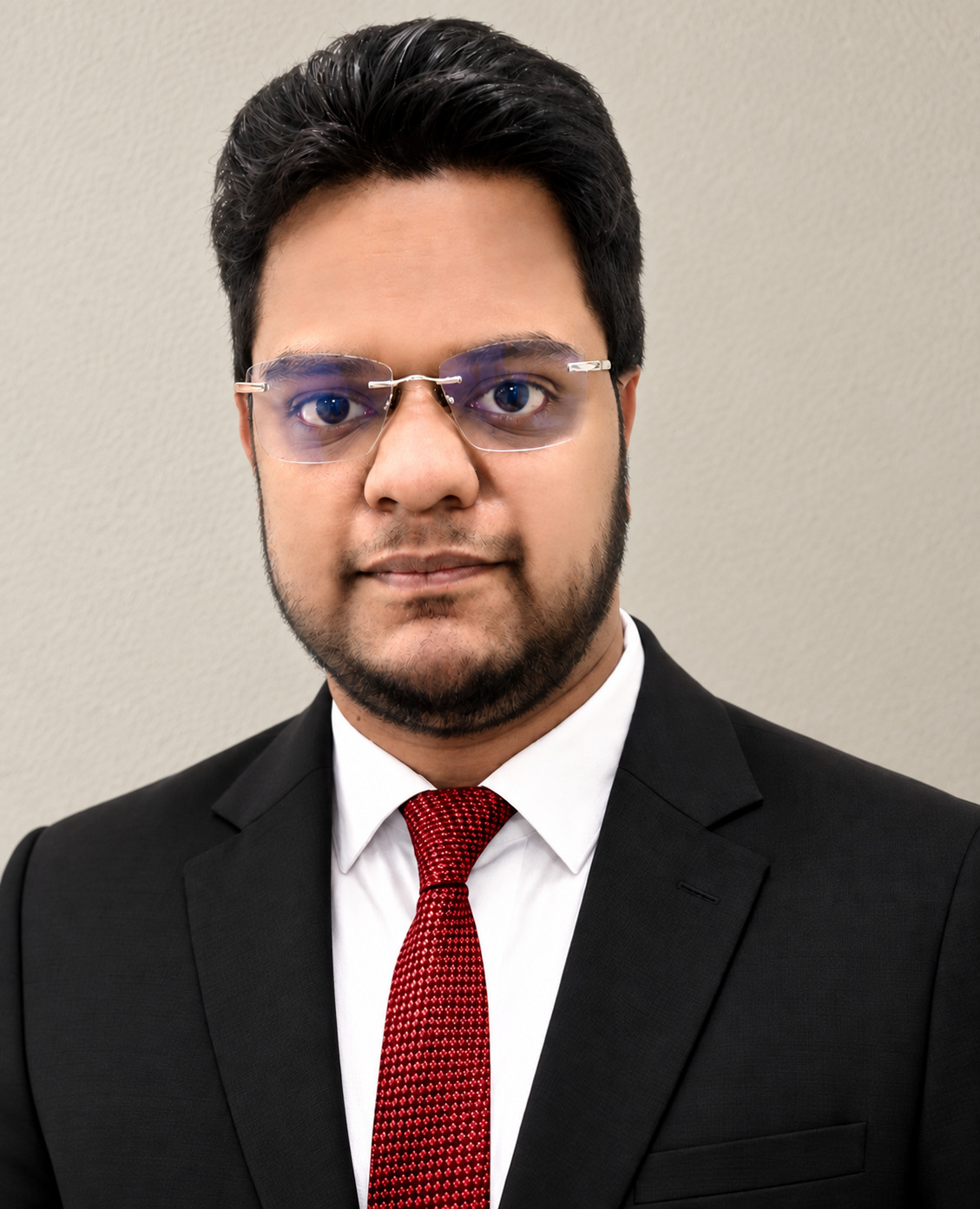}}]{Gaurav K. Pandey} (Member, IEEE)
received the Ph.D. degree in Electronics and Communication Engineering from the National Institute of Technology Silchar, India, in 2025, and the M.Tech. degree in Communication Engineering from the National Institute of Technology Nagaland, India, in 2018. He was a recipient of the Government of India fellowships for pursuing his doctoral and master's studies through the GATE. He has received various Best Paper Awards at international conferences and is an inventor of two Indian patents.

Dr. Pandey is currently a Researcher with the Faculty of Electrical Engineering and Informatics, Technical University of Košice (TUKE), Slovakia. His research interests include non-terrestrial networks, reconfigurable intelligent surfaces, physical-layer security, semantic communications, and integrated sensing and communications. His current research focuses on deep reinforcement learning, quantum deep reinforcement learning, and intelligent resource allocation for AI-native 6G networks. He has served as a reviewer for more than 50 IEEE transactions and conferences. 

\end{IEEEbiography}

\begin{IEEEbiography}
[{\includegraphics[width=1in,height=1.25in,clip,keepaspectratio]{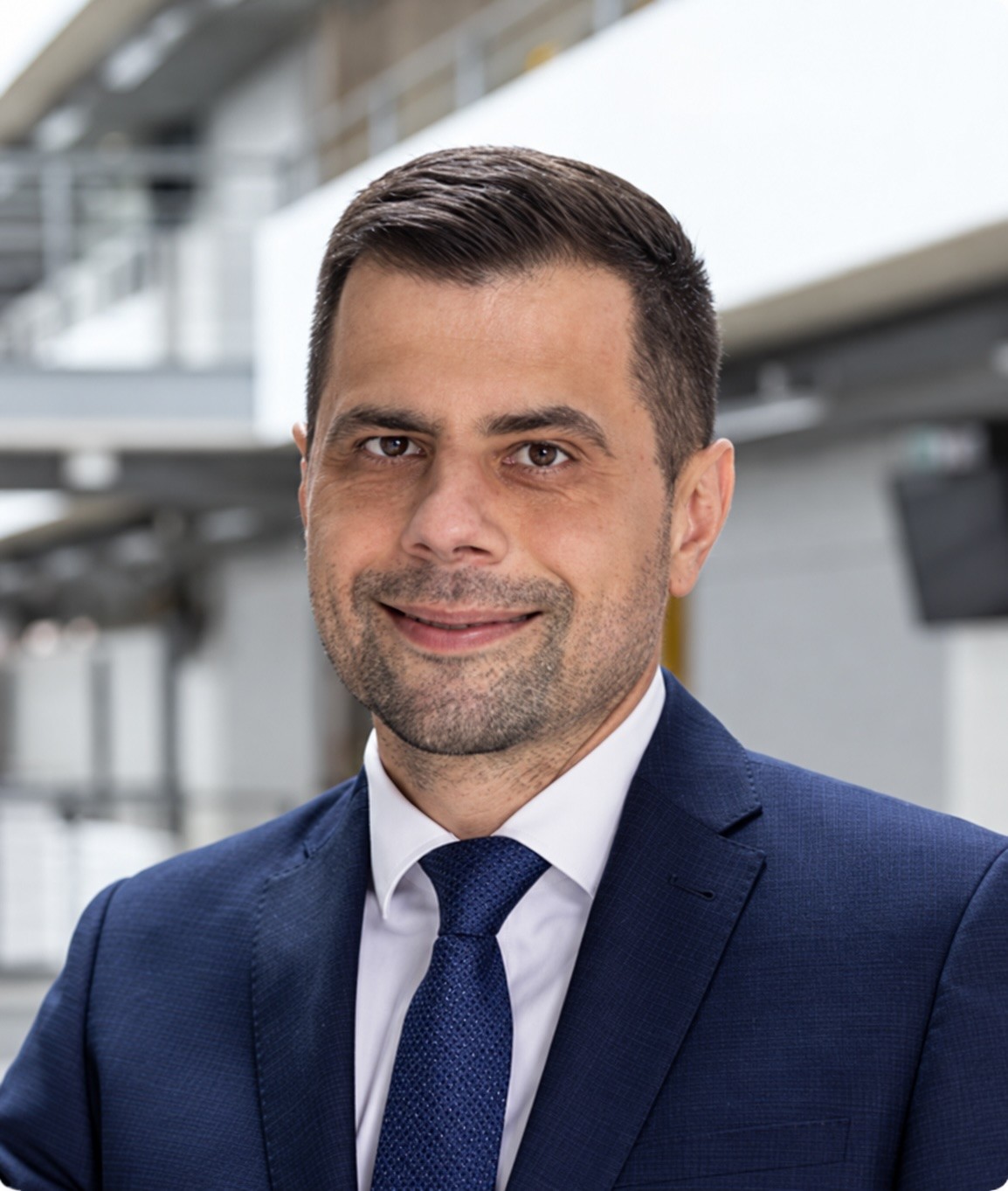}}]
{Juraj Gazda} is currently a Vice-Rector for Innovation and Technology Transfer and Professor with the Faculty of Electrical Engineering at the Technical University of Košice (TUKE), Slovakia. 

He has been a guest researcher at Ramon Llull University, Barcelona, and the Technical University of Hamburg-Harburg, and has been involved in development projects for Nokia Siemens Networks (NSN) and Ericsson. In 2017, he was recognized as the Best Young Scientist at TUKE.

Prof. Juraj Gazda currently serves on the executive board of IT Valley, an innovation ecosystem supporting collaboration between academia, industry, and public sector, and on the executive board of AI4Slovakia, a national initiative driving the adoption and strategic development of artificial intelligence in Slovakia.
He is also an editor of the KSII Transactions on Internet and Information Systems and a guest editor for Wireless Communications and Mobile Computing (Wiley).
\end{IEEEbiography}

\end{document}